\documentclass[unnumsec,namedate,webpdf,traditional,medium]{oup-authoring-template}

\onecolumn 

\graphicspath{{Fig/}}

\theoremstyle{thmstyleone}%
\newtheorem{theorem}{Theorem}
\theoremstyle{thmstyletwo}%
\theoremstyle{thmstylethree}%
\newtheorem{definition}{Definition}
\usepackage{rotating}
\usepackage{bm}
\usepackage{natbib}
\usepackage{url}
\usepackage{bm}
\usepackage{amsthm,amsmath,amsfonts,amssymb}
\usepackage{rotating}
\DeclareMathOperator*{\median}{median}
\newtheorem{condition}{Condition}
\newtheorem{lemma}{Lemma}
\begin{document}

\journaltitle{}
\DOI{DOI HERE}
\copyrightyear{2022}
\pubyear{2019}
\access{Advance Access Publication Date: Day Month Year}
\appnotes{Paper}

\firstpage{1}


\title[EGG: Estimation of Genetic Graph]{Estimation of the genetic Gaussian network using GWAS summary data}

\author[1]{Yihe Yang \ORCID{0000-0001-6563-3579}}
\author[1]{Noah Lorincz-Comi \ORCID{0000-0002-0517-2499}}
\author[1,$\ast$]{Xiaofeng Zhu \ORCID{0000-0003-0037-411X}}

\authormark{Yang et al.}

\address[1]{\orgdiv{Department of Population and Quantitative Health Sciences, School of Medicine}, \orgname{Case Western Reserve University}, \orgaddress{\street{10900 Euclid Ave, Cleveland}, \postcode{44106}, \state{OH}, \country{USA}}}

\corresp[$\ast$]{Corresponding author. \href{email:xxz10@case.edu}{xxz10@case.edu}}

\received{Date}{0}{Year}
\revised{Date}{0}{Year}
\accepted{Date}{0}{Year}



\abstract{Genetic Gaussian network of multiple phenotypes constructed through the genetic correlation matrix is informative for understanding their biological dependencies. However, its interpretation may be challenging because the estimated genetic correlations are biased due to estimation errors and horizontal pleiotropy inherent in GWAS summary statistics. Here we introduce a novel approach called Estimation of Genetic Graph (EGG), which eliminates the estimation error bias and horizontal pleiotropy bias with the same techniques used in multivariable Mendelian randomization. The genetic network estimated by EGG can be interpreted as representing shared common biological contributions between phenotypes, conditional on others, and even as indicating the causal contributions. We use both simulations and real data to demonstrate the superior efficacy of our novel method in comparison with the traditional network estimators.  R package \texttt{EGG} is available on \url{https://github.com/harryyiheyang/EGG}.}
\maketitle

\section{Introduction}
Gaussian graphical model (GGM) is one of the most frequently used models to quantify and visualize the dependence structure of multiple phenotypes via an undirected graph/network \citep{lauritzen1996graphical}. In GGM, the network of multiple Gaussian variables can be described by either their precision matrix or their partial correlation coefficients. This results in two primary schemes that have been proposed for estimating the network of Gaussian variables: one scheme estimates the precision matrix through maximum likelihood estimation (MLE) of multivariate Gaussian distribution \citep{yuan2007model}, while the other estimates the partial correlation coefficients by solving node-wise lasso regression \citep{meinshausen2006high}. Numerous methods have extended the GGM to describe the graph of non-Gaussian continuous  \citep{ravikumar2011high} and discrete variables \citep{ravikumar2010high}.

Despite its widespread application, GGM faces several challenges. Firstly, the network estimates can be biased due to unobserved confounders. Indeed, GGM measures the partial correlation coefficient between variable pairs, conditioned on all other variables. If the confounders affecting these variables are unobserved, this could theoretically cause incorrect identification of significant partial correlation and a biased network estimate \citep{buhlmann2011statistics}. Second, traditional statistical methods, including GGM, often require comprehensive datasets that encompass all observed phenotypes of interest at the individual level. This reliance limits their effectiveness in scenarios with smaller, more fragmented samples, typical in fields like medicine where data is collected across various cohorts focusing on specific phenotypes \citep{le2020challenges}. 

Genome-wide association study (GWAS) provides a new opportunity to model the biological relationship of multiple phenotypes without individual-level data. Specifically, GWAS estimates the associations of a phenotype with genetic variants, or single nucleotide polymorphisms (SNPs), across the genome. GWAS summary data typically includes the estimated genetic effect sizes, the corresponding standard errors (SEs), P-values, allele frequencies, and qualities, and are usually accessible in public databases and websites such as dbGaP \citep{mailman2007ncbi} and GWAS Catalog \citep{macarthur2017new}. Meta-analyses of GWAS summary data from multiple cohorts have markedly improved the precision of genetic association estimates because of the substantial sample size increment, even reaching over one million for some phenotypes \citep{graham2021power}. In addition, since the genotypes of individuals are randomly inherited from their parents and generally do not change during their lifetime, genetic variants are supposedly independent of underlying confounders and hence the inferences made by GWAS summary data are considered to be robust against reverse causation and confounder bias \citep{bulik2015atlas}. Epidemic studies with GWAS summary data would supposedly yield more reliable results compared to those relying on individual-level data \citep{abdellaoui202315}.

\citet{bulik2015ld,bulik2015atlas} introduced linkage disequilibrium score regression (LDSC), a widely used for estimating narrow-sense heritability and genetic correlations between traits based on GWAS summary statistics. Subsequently, \citet{shi2017local,zhao2022genetic,wang2022estimation} developed multiple alternatives to estimate genetic correlations under the fixed effect model of genetic effect sizes. \citet{yan2020fam222a} further demonstrated the detection of disease-associated genes with GWAS summary statistics by employing the genetic correlation estimation and weighted gene co-expression network analysis (WGCNA) \citep{langfelder2008wgcna}. However, while genetic correlations offer insights into shared genetic contributions between traits, they do not imply causality. Confounding factors, genetic pleiotropy, and measurement error bias challenge their interpretation. On the other side, Mendelian randomization (MR) has been used to infer causal relationships among phenotypes using genetic instrument variables \citep{burgess2013mendelian}, and has recently been extended to search for gene-environment interactions \citep{zhu2023new}. Although MR has been extended to multiple exposures (MVMR), it does not investigate the relationship within exposures \citep{sanderson2019examination}. \citet{lin2023combining} proposed a new approach to investigate the potential causal diagram within multiple phenotypes, beginning with bidirectional MR \citep{welsh2010unraveling} and subsequently applying network deconvolution \citep{feizi2013network}. Whereas, the bias inherited for MR will also be carried forward to this network deconvolution approach, including weak instrument bias and horizontal pleiotropy bias \citep{lorincz-comi2022mrbee}. Additionally, the mathematics used in the network deconvolution has been criticized \citep{Pachter2014NetworkNonsense}. To the best of our knowledge, there has not yet been a method for estimating the Gaussian network of multiple phenotypes directly from GWAS summary statistics. Since the Gaussian network infers conditional correlations with well-established statistical properties, extending it to GWAS summary statistics offers a robust and statistically sound framework for understanding biological interdependencies among phenotypes.

In this paper, we develop a novel method named Estimation of Genetic Graph (EGG), which estimates the Gaussian network of multiple phenotypes using their GWAS summary statistics. The inferred network can be interpreted as representing the shared common genetic contributions among phenotypes conditional on others, and potentially as indicating the causal contributions. In particular, EGG addresses two specific features of GWAS summary statistics, estimation errors of genetic effect sizes \citep{ye2021debiased} and horizontal pleiotropy \citep{zhu2021iterative}, by using the same techniques used in MVMR analysis \citep{bowden2016consistent,lorincz-comi2022mrbee}. This distinguishes EGG from the existing methods of GGM such as graphical lasso \citep{friedman2008sparse} and neighborhood selection \citep{meinshausen2006high}. We used both simulations and real data to demonstrate the superior efficacy of our proposed method in comparison with the traditional graph estimators. We applied EGG to analyze 20 phenotypes in the European (EUR) and East Asian (EAS) populations, respectively, including coronary artery disease (CAD) \citep{aragam2022discovery}, type 2 diabetes (T2D) \citep{vujkovic2020discovery}, ischemic stroke \citep{mishra2022stroke}, etc. The inferred genetic network offers novel insight into the causal pathways of complex metabolic and cardiovascular disease.
\section{Preliminary}
In this section, we introduce three basic concepts: 1) GGM, 2) random effect model, and 3) GWAS. We will also illustrate the estimation errors of genetic effect sizes and horizontal pleiotropy, the two specific features of GWAS summary data.
\subsection{Notation}
For a vector $\boldsymbol a=(a_1,\dots,a_p)^\top$, $||\boldsymbol a||_q=(\sum_{j=1}^p|a|_j^q)^{1/q}$. For a $(p\times p)$ symmetric matrix $\mathbf A$, $\sigma_{\max }(\mathbf A)$ and $\sigma_{\min }(\mathbf A)$ denote its maximum/minimum eigenvalues, $\mathbf A=\sum_{j=1}^p\sigma_j(\mathbf A)\boldsymbol U_j\boldsymbol U_j^\top$ represents its eigenvalue decomposition, $f(\mathbf A)=\sum_{j=1}^pf(\sigma_j(\mathbf A))\boldsymbol U_j\boldsymbol U_j^\top$, and $[\mathbf A,\delta]_+=\sum_{j=1}^p\max(\sigma_j(\mathbf A),\delta)\boldsymbol U_j\boldsymbol U_j^\top$.  For a general matrix $\mathbf A=(A_{ij})_{m\times p}$, $\|\mathbf A\|_\texttt{F}=\{\texttt{tr}(\mathbf A^\top\mathbf A)\}^{\frac12}$ and $||\mathbf A||_q=\sup_{||\boldsymbol a||_q=1}||\mathbf A\boldsymbol a||_q$. In the special cases $q=1,2,\infty$, $||\mathbf A||_1=\max_{1\leq j\leq p}\sum_{i=1}^m|A_{ij}|$, $||\mathbf A||_2=\sigma_{\max }(\mathbf A^\top\mathbf A)^{1/2}$, and $||\mathbf A||_\infty=\max_{1\leq i\leq m}\sum_{j=1}^p|A_{ij}|$. $\texttt{I}(\cdot)$ is the indicator function, $\texttt{diag}(\cdot)$ converts a vector to a diagonal matrix, and $\texttt{cov2cor}(\cdot)$ converts a covariance matrix to a correlation matrix. Additionally, under general circumstances, $i$ denotes the index for individuals, $j$ and $t$ indicate the indices for genetic variants, and $k$ and $s$ refer to the indices for phenotypes.
\subsection{Gaussian Graphical Model}
A graph or network, denoted as $\mathcal G$, is composed of a set of vertices $\mathcal V$ and a set of edges $\mathcal E\subset\mathcal V\times \mathcal V$. An edge is classified as undirected if both $(k,s)$ and $(s, k)$ belong to $\mathcal E$. Conversely, an edge is considered directed from vertex $j$ to vertex $k$ if $(k,s)$ is in $\mathcal E$ but $(s, k)$ is not. In a probabilistic graphical model, the vertices of a graph correspond to a collection of random variables $\boldsymbol X_i=(X_{i1},\dots,X_{ip})^\top\sim\mathcal P$ where $\mathcal V = \{1,\dots,p\}$ and $\mathcal P$ is the probability distribution of $X_i$. In addition, if $\mathcal G$ is an undirected graph, then there is a global Markov property: $X_{ik}\perp X_{is}|\boldsymbol X_{i}^{\mathcal V/{k,s}}$ if and only if the pair of unconnected vertices $(k,s)\not\in\mathcal E$,  where $\boldsymbol X_{i}^{\mathcal V/{k,s}}$ is a sub-vector of $\boldsymbol X_i$ excluding $X_{ik}$ and $X_{ik}$. In other words, two variables are independent conditional on the other variables is equivalent to stating that they are not connected by an edge in an undirected graph.

The GGM represents conditionally independent relationships in a multivariate Gaussian distribution using an undirected graph/network. Let $\boldsymbol X_i$ be a $p$-variate Gaussian variable with a covariance matrix $\mathbf\Sigma$. The precision matrix of $\boldsymbol X_i$ is denoted as $\mathbf\Theta=\mathbf\Sigma^{-1}$. Due to the Gaussianity, the following equivalences hold:
\begin{align}
(k,s)\text{ and }(s,k)\not\in\mathcal E\Leftrightarrow X_{ik}\perp X_{is}|\boldsymbol X_{i}^{\mathcal V/{k,s}}\Leftrightarrow\Theta_{ks}=\Theta_{sk}=0.
\label{equiv1}
\end{align}
It allows us to infer which vertices are connected by edges by investigating which entries in $\mathbf\Theta$ are non-zero. An alternative to characterizing a Gaussian network involves using regression coefficients or partial correlation coefficients \citep{meinshausen2006high}. Refer to \citet{lauritzen1996graphical} and \citet{buhlmann2011statistics} for details.

\subsection{Random Effect Model}
Consider a vector of multiple phenotypes $\boldsymbol X_i=(X_{i1},\dots,X_{ip})^\top$ of an individual $i$ with $X_{ik}=\eta_{ik}+\epsilon_{ik}$,
where $\eta_{ik}$ is determined by an individual's genotypes and $\epsilon_{ik}$ is a non-genetic effect orthogonal to the genetic effect $\eta_{ik}$. By separating $X_{ik}$ into $\eta_{ik}$ and $\epsilon_{ik}$, the covariance between $X_{ik}$ and $X_{is}$ is then 
\begin{align}
\texttt{cov}(X_{ik},X_{is})=\texttt{cov}(\eta_{ik},\eta_{is})+\texttt{cov}(\epsilon_{ik},\epsilon_{is}),
\end{align}
where $\texttt{cov}(\eta_{ik},\eta_{is})$ is called the genetic covariance and $\texttt{cov}(\epsilon_{ik},\epsilon_{is})$ is considered as the covariance of non-genetic effect. In addition, the narrow-sense heritability, the fraction of phenotypic variance that can be attributed to the additive effects of variants, is defined as $h_k^2=\texttt{var}(\eta_{ik})/\texttt{var}(X_{ik})$.

The random effect model \citep{yang2010common} assumes 
\begin{align}
\eta_{ik}=\sum_{j=1}^mG_{ij}\beta_{jk},
\end{align}
where $G_{i1},\dots,G_{im}$ are $m$ genetic variants, $\beta_{j1},\dots,\beta_{jp}^\top$ are $m$ genetic effects on the $k$th trait, and $\{G_{ij}\}$ and $\{\beta_{jk}\}$ are mutually independent. In random effect model, the genetic effects of a variant for multiple phenotypes $\boldsymbol\beta_j=(\beta_{j1},\dots,\beta_{jp})^\top$ are assumed to follow 
\begin{align}
\boldsymbol\beta_j\sim\mathcal{N}(\mathbf0,\mathbf\Sigma_{\boldsymbol\beta}),
\end{align}
where $\mathbf\Sigma_{\boldsymbol\beta}$ is called as the genetic covariance matrix with the $k$th diagonal entry being $h_k^2/m$.
As for $\boldsymbol G_i=(G_{i1},\dots,G_{im})^\top$, $\texttt{E}(\boldsymbol G_i)=\mathbf0$ and $\texttt{cov}(\boldsymbol G_i)=\mathbf{L}$ which is known as the LD matrix with diagonal entries being one. Under this setting, the genetic covariance
\begin{align}
\texttt{cov}(\eta_{is},\eta_{ik})=\texttt{E}\bigg\{\bigg(\sum_{j=1}^mG_{ij}\beta_{jk}\bigg)\bigg(\sum_{t=1}^mG_{it}\beta_{tk}\bigg)\bigg\}=\sum_{j=1}^m\bigg(\texttt{E}(G_{ij}^2)\texttt{cov}(\beta_{jk},\beta_{js})\bigg)=m\Sigma_{\beta_k\beta_s},
\end{align}
which is obtained by using the condition $\texttt{E}(G_{ij}\beta_{jk}G_{it}\beta_{tk})=\texttt{E}(\beta_{jk}\beta_{tk})\texttt{E}(G_{ij}G_{it})=0$ since $\texttt{E}(\beta_{jk}\beta_{tk})=0$ if $j\neq t$. In other words, the genetic covariance between two traits are indeed the cumulative covariance between their genetic effects. In addition, it is widely applied to simplify the random effect model by considering only one independent variant in each LD region, resulting in $\mathbf{L}=\mathbf I_m$.  In practice, the clumping plus thresholding (C+T) method can acquire independent variants in each LD region \citep{purcell2007plink}. 
\subsection{Genome-wide Association Studies}
GWAS usually refers to the study that estimates $\beta_{jk}$ by
\begin{align}
\hat\beta_{jk}=\frac1{n_k}\sum_{i=1}^{n_k}X_{ik}G_{ij},
\end{align}
where $n_k$ is the sample size of the GWAS of $X_{ik}$. The GWAS summary data usually contain the effect estimate $\hat\beta_{jk}$, its SE and P-value, and other information of the $j$th variant. In the field of GWAS, different phenotypes are often studied across various cohorts, and multiple cohorts may share partial common samples. Existing methods based on GWAS summary data treat the effects of individual genetic variants analogously to individual subjects in traditional studies, allowing investigating multiple phenotypes without individual-level data \citep{bulik2015ld,bulik2015atlas,ruan2022improving,lorincz-comi2022mrbee}.

However, the estimation error in the GWAS effect size estimate may introduce bias into the current network methods. Under the condition $\texttt{cov}(\boldsymbol G_i)=\mathbf I_m$, \citet{lorincz-comi2022mrbee}
showed\begin{align}
\hat\beta_{jk}=\beta_{jk}+\omega_{jk},
\end{align}
where $\omega_{jk}=\sum_{i}G_{ij}(X_{ik}-G_{ij}\beta_{jk})/n_{k}$. While the GWAS effect estimates $\hat\beta_{jk}$ is unbiased for the true effect $\beta_{jk}$, the genetic covariance estimate based on GWAS effect estimates is biased due to their estimation errors:
\begin{align}
\texttt{cov}(\hat\beta_{jk},\hat\beta_{js})=\texttt{cov}(\beta_{jk},\beta_{js})+\texttt{cov}(\omega_{jk},\omega_{js}),
\end{align}
where \citet{lorincz-comi2022mrbee}  proved 
\begin{align}
\texttt{cov}(\omega_{jk},\omega_{js})\approx \frac{n_{ks}}{n_kn_s}\texttt{cov}(X_{ik},X_{is})
\end{align}
with $n_{ks}$ being the overlapping sample size between the GWAS of $X_{ik}$ and $X_{is}$. As a result, the confounders in $X_{ik}$ and $X_{is}$ can bias the genetic covariance estimate, making the GGM methods that use a correlation matrix estimate as the input unreliable.

Horizontal pleiotropy or horizontal pleiotropic variants refer to the variants that are associated with more than two different traits \citep{zhu2021iterative}. Statistically, we can differentiate horizontal pleiotropic variants from the rest variants by a mixture of Gaussian distribution:
\begin{align}
\beta_{jk}\sim\pi^\texttt{pleio}
\mathcal{N}(\gamma_{jk},\Sigma_{\beta_k\beta_k})+(1-\pi^\texttt{pleio})\mathcal{N}(0,\Sigma_{\beta_k\beta_k}),
\end{align}
where $\gamma_{jk}$ is a constant used to describe the pleiotropic effect and $\pi^\texttt{pleio}$ is the fraction of pleiotropic variants. In practice, the fraction $\pi^\texttt{pleio}$ is usually small, and hence the variants with pleiotropic effects can also be viewed as outliers \citep{bowden2015mendelian}. An alternative distribution to describe the horizontal pleiotropy is 
\begin{align}
\beta_{jk}\sim\mathcal{F}(0,\Sigma_{\beta_k\beta_k})
\end{align}
where $\mathcal{F}(\mu,\sigma^2)$ is a distribution with mean $\mu$ and variance $\sigma^2$. This distribution usually has heavier tails than the Gaussian distribution (e.g., the Student's t distribution), and describes the pleiotropy as heavy-tail errors.  In MR, how to address the bias caused by horizontal pleiotropy is a challenging issue. The solutions include 1) detecting the pleiotropy using hypothesis test \citep{zhu2021iterative}, 2) reducing the pleiotropic effect with robust tool \citep{bowden2015mendelian}, and 3) estimating causal and pleiotropy effects by Bayesian mixture model \citep{morrison2020mendelian}. However, the existing genetic correlation methods \citep{bulik2015atlas,zhao2022genetic,wang2022estimation} have not investigated the pleiotropy bias. 
\section{Estimation of Genetic Graph}
In this paper, our goal is to estimate the precision matrix $\mathbf\Theta_{\boldsymbol\beta}=\mathbf\Sigma_{\boldsymbol\beta}^{-1}$, which can provide an effective and visual way to understand the dependence skeleton of $\boldsymbol X_j$. We first resolve the estimation error and pleiotropy bias, next provide a new algorithm to estimate the genetic precision matrix, and finally investigate the convergence rate of the network estimate.
\subsection{Estimation of Estimation Error Covariance Matrix}
Let $\hat{\boldsymbol\beta}_j=\boldsymbol\beta_j+\boldsymbol \omega_j$, where $\boldsymbol\beta_j$ is the true genetic effect of the $j$th variant, $\hat{\boldsymbol\beta}_j$ is its GWAS marginal association estimate, and $\boldsymbol \omega_j$ is the estimation error. The covariance matrices are:
\begin{align}
\texttt{cov}(\hat{\boldsymbol\beta}_j)=\mathbf\Sigma_{\hat{\boldsymbol\beta}},\quad \texttt{cov}(\boldsymbol\beta_j)=\mathbf\Sigma_{\boldsymbol\beta},\quad\texttt{cov}(\boldsymbol \omega_j)=\mathbf\Sigma_{\boldsymbol \omega},
\end{align}
and $\Sigma_{\hat\beta_k\hat\beta_s}$, $\Sigma_{\beta_k\beta_s}$, and $\Sigma_{\omega_k\omega_s}$ are the $(k,s)$th entries in them. 
If the covariance matrix of estimation error $\mathbf\Sigma_{\boldsymbol \omega}$ is estimable, we can unbiasedly estimate the true genetic covariance matrix $\mathbf\Sigma_{\boldsymbol\beta}$ by:
\begin{align}
\widehat{\mathbf\Sigma}_{\boldsymbol\beta}=\frac1m\sum_{j=1}^m\hat{\boldsymbol\beta}_j\hat{\boldsymbol\beta}_j^\top-\widehat{\mathbf\Sigma}_{\boldsymbol \omega},
\label{Pearsoncov}
\end{align}
where $\widehat{\mathbf\Sigma}_{\boldsymbol \omega}$ is the unbiased estimate of $\mathbf\Sigma_{\boldsymbol \omega}$. We call the covariance matrix estimated by (\ref{Pearsoncov}) the Pearson's r covariance estimate $\widehat{\mathbf\Sigma}_{\boldsymbol\beta}^\texttt{Pearson}$.

In the literature, two methods are commonly used to estimate the covariance matrix of estimation error $\mathbf\Sigma_{\boldsymbol\omega}$: LDSC \citep{bulik2015atlas} and null effect estimate \citep{zhu2015meta}. LDSC estimates $\mathbf\Sigma_{\boldsymbol\beta}$ and $\mathbf\Sigma_{\boldsymbol \omega}$ entry-by-entry, often resulting in non-positive definite estimates of covariance matrices. In contrast, estimating $\mathbf\Sigma_{\boldsymbol\omega}$ from the insignificant effect estimates is more straightforward and enjoys greater computational efficiency than LDSC. Let $F_{i1},\dots,F_{iM}$ be $M$ independent genetic variants that are not associated with a trait, and let $\hat b_{jk}=n_k^{-1}\sum_{i=1}^{n_k}F_{ij}X_{ik}$ be the insignificant effect estimate. \citet{lorincz-comi2022mrbee} proved that $\hat b_{jk}$ and $\omega_{jk}$ has the same asymptotic distribution, allowing estimating $\mathbf\Sigma_{\boldsymbol \omega}$ by
\begin{align}
\widehat{\mathbf\Sigma}_{\boldsymbol \omega}=\frac1M\sum_{j=1}^M\hat{\boldsymbol b}_j\hat{\boldsymbol b}_j^\top,
\label{Pearson}
\end{align}
where $\hat{\boldsymbol b}_j=(\hat b_{jk},\dots,\hat b_{jp})^\top$.   In practice, there are millions of common variants across the genome, but only a small fraction of these are significantly associated with a trait. Therefore, it is feasible to obtain a substantial number of insignificant variants and $\widehat{\mathbf\Sigma}_{\boldsymbol\omega}$ can be estimated precisely.
\subsection{Robust Estimation of Genetic Covariance Matrix}
We propose the rank-based covariance matrix estimate as an alternative to $\widehat{\mathbf\Sigma}_{\boldsymbol\beta}^\texttt{Pearson}$, which has been verified robust to potential outliers and heavy-tail errors \citep{avella2018robust}. As the horizontal pleiotropy have similar performanes as outliers, this rank-based covariance matrix is supposedly more accurate than $\widehat{\mathbf\Sigma}_{\boldsymbol\beta}^\texttt{Pearson}$. 

Specifically, we employ two robust methods to separately estimate the correlation matrix and the diagonal standard deviation matrix of $\hat{\boldsymbol\beta}$, denoted as $\widehat{\mathbf R}_{\hat{\boldsymbol\beta}}$ and $\widehat{\mathbf D}_{\hat{\boldsymbol\beta}}$, respectively. For the correlation matrix, we consider the following Spearman's rho correlation:
\begin{align}
\hat\rho_{ks}=\frac{\sum_{j=1}^m(\hat r_{k}^j-\frac{m+1}{2})(\hat r_{s}^j-\frac{m+1}{2})}{\sqrt{\sum_{j=1}^m(\hat r_{k}^j-\frac{m+1}{2})^2}\sqrt{\sum_{j=1}^m(\hat r_{s}^j-\frac{m+1}{2})^2}},
\end{align}
where $\hat r^j_k$ represents the rank of $\hat\beta_{jk}$ among $\hat\beta_{1k},\dots,\hat\beta_{mk}$. According to \citet{avella2018robust}, we then recover original correlation matrix of $\hat{\boldsymbol\beta}_j$ by:
\begin{align}
\hat R_{\beta_k\beta_s}^\texttt{Spearman}=2\sin(\pi\hat\rho_{ks}/6),
\end{align}
and let $\widehat{\mathbf R}_{\boldsymbol\beta}^\texttt{Spearman}=(\hat R_{\beta_k\beta_s}^\texttt{Spearman})_{p\times p}$. On the other hand, we use the median
absolute deviation (MAD) as the robust standard deviation estimate:
\begin{align}
\hat D_{\hat\beta_k}^\texttt{MAD}=1.483\median_{1\leq j\leq m}(|\hat\beta_{jk}-\median_{1\leq j\leq m}(\hat\beta_{jk})|),
\end{align}
and let $\widehat{\mathbf D}_{\hat{\boldsymbol\beta}}^\texttt{MAD}=\texttt{diag}(\hat D_{\hat\beta_1}^\texttt{MAD},\dots,\hat D_{\hat\beta_p}^\texttt{MAD})$. This results in a robust estimate of $\mathbf\Sigma_{\hat{\boldsymbol\beta}}$ given by:
\begin{align}
\widehat{\mathbf\Sigma}_{\boldsymbol\beta}^\texttt{Spearman}=\widehat{\mathbf D}_{\hat{\boldsymbol\beta}}^\texttt{MAD}\widehat{\mathbf R}_{\hat{\boldsymbol\beta}}^\texttt{Spearman}\widehat{\mathbf D}_{\hat{\boldsymbol\beta}}^\texttt{MAD}-\widehat{\mathbf\Sigma}_{\boldsymbol \omega},
\label{Spearmancov}
\end{align}
which is called Spearman's rho covariance estimate.  Other robust covariance matrix estimates such as those based on Kendall's tau correlation are also commonly used in practice, whose asymptotic properties are similar to Spearman's rho correlation \citep{avella2018robust}. Here, we use Spearman's rho correlation as a representative of these robust estimators.
\subsection{Estimation of Genetic Precision Matrix}
Let $\mathbf\Theta_{\boldsymbol\beta}=\mathbf\Sigma_{\boldsymbol\beta}^{-1}$, which is the genetic precision matrix to be estimated. We propose to unbiasedly estimate it through the following constrained minimization:
\begin{align}
\widehat{\mathbf\Theta}_{\boldsymbol\beta}=\arg\min_{\mathbf\Theta_{\boldsymbol\beta}}\bigg\{\texttt{entropy}(\widehat{\mathbf\Sigma}_{\boldsymbol\beta},\mathbf\Theta_{\boldsymbol\beta})+\sum_{1\leq j<j'\leq p}P_\lambda(\Theta_{\beta_j\beta_{j'}})\bigg\},
\label{penalizedentropy}
\end{align}
subject to $\sigma_\texttt{min}(\mathbf\Theta_{\boldsymbol\beta})>\delta$, where the entropy loss function \citep{yang2021estimation} is:
\begin{align}
\texttt{entropy}(\widehat{\mathbf\Sigma}_{\boldsymbol\beta},\mathbf\Theta_{\boldsymbol\beta})=\texttt{tr}(\widehat{\mathbf\Sigma}_{\boldsymbol\beta}\mathbf\Theta_{\boldsymbol\beta})-\log\det(\widehat{\mathbf\Sigma}_{\boldsymbol\beta}\mathbf\Theta_{\boldsymbol\beta})-p,
\end{align}
$\widehat{\mathbf\Sigma}_{\boldsymbol\beta}$ can be the Pearson's r estimate (\ref{Pearsoncov}) or the Spearman's rho estimate (\ref{Spearmancov}), $P_\lambda(\cdot)$ is a non-convex penalty with a tuning parameter $\lambda$, and $\delta>0$ is a given threshold. In this constrained minimization, $P_\lambda(\cdot)$ is used to select the non-zero entries in $\mathbf\Theta_{\boldsymbol\beta}$ \citep{tibshirani1996regression}, and $\sigma_\texttt{min}(\mathbf\Theta_{\boldsymbol\beta})>\delta$ is applied to guarantee that $\widehat{\mathbf\Theta}_{\boldsymbol\beta}$ is positive definite \citep{zhang2014sparse}.

We apply the ADMM algorithm \citep{boyd2011distributed} to solve it distributively. The ADMM algorithm first converts  (\ref{penalizedentropy}) into the minimization below:
\begin{align}
\widehat{\mathbf\Theta}_{\boldsymbol\beta},\widehat{\mathbf\Omega}_{\boldsymbol\beta},\widehat{\mathbf\Gamma}_{\boldsymbol\beta}=\arg\min_{\mathbf\Theta_{\boldsymbol\beta},\mathbf\Omega_{\boldsymbol\beta},\mathbf\Gamma_{\boldsymbol\beta}}\bigg\{
\texttt{entropy}(\widehat{\mathbf\Sigma}_{\boldsymbol\beta},\mathbf\Theta_{\boldsymbol\beta})+\sum_{1\leq j<j'\leq p}P_\lambda(\Omega_{\beta_j\beta_{j'}}),
\bigg\}
\label{admm1}
\end{align}
subject to $\mathbf\Theta_{\boldsymbol\beta}=\mathbf\Omega_{\boldsymbol\beta}$, $\mathbf\Theta_{\boldsymbol\beta}=\mathbf\Gamma_{\boldsymbol\beta}$, $\sigma_\texttt{min}(\mathbf\Gamma_{\boldsymbol\beta})>\delta$. The ADMM algorithm then considers the following Lagrange augmented function of (\ref{admm1}):
\begin{align}
\mathcal Q(\mathbf\Theta_{\boldsymbol\beta},&\mathbf\Omega_{\boldsymbol\beta},\mathbf\Gamma_{\boldsymbol\beta},\mathbf\Lambda_1,\mathbf\Lambda_2)=\texttt{entropy}(\widehat{\mathbf\Sigma}_{\boldsymbol\beta},\mathbf\Theta_{\boldsymbol\beta})+\sum_{1\leq j<j'\leq p}P_\lambda(\Omega_{\beta_j\beta_{j'}})\notag\\
&+\texttt{tr}(\mathbf\Lambda_1(\mathbf\Theta_{\boldsymbol\beta}-\mathbf\Omega_{\boldsymbol\beta}))
+\frac\psi2||\mathbf\Theta_{\boldsymbol\beta}-\mathbf\Omega_{\boldsymbol\beta}||_\texttt{F}^2+\texttt{tr}(\mathbf\Lambda_2(\mathbf\Theta_{\boldsymbol\beta}-\mathbf\Gamma_{\boldsymbol\beta}))+\frac\psi2||\mathbf\Theta_{\boldsymbol\beta}-\mathbf\Gamma_{\boldsymbol\beta}||_\texttt{F}^2,
\end{align}
subject to $\sigma_\texttt{min}(\mathbf\Gamma_{\boldsymbol\beta})>\delta$. Here $\mathbf\Lambda_1$ and $\mathbf\Lambda_2$ are the Lagrange multipliers corresponding to constraints  $\mathbf\Theta_{\boldsymbol\beta}=\mathbf\Omega_{\boldsymbol\beta}$ and $\mathbf\Theta_{\boldsymbol\beta}=\mathbf\Gamma_{\boldsymbol\beta}$, and the quadratic terms $\frac\psi2||\mathbf\Theta_{\boldsymbol\beta}-\mathbf\Gamma_{\boldsymbol\beta}||_\texttt{F}^2$ and $\frac\psi2||\mathbf\Theta_{\boldsymbol\beta}-\mathbf\Gamma_{\boldsymbol\beta}||_\texttt{F}^2$ with a tuning parameter $\psi$ are imposed to smooth the constraints. 

As the largest advantage, the ADMM algorithm only minimizes a parameter given other parameter estimates at a time, and thus divides a complex minimization into a series of simpler ones. The update of $\mathbf\Theta_{\boldsymbol\beta}$ is 
\begin{align}
\mathbf\Theta_{\boldsymbol\beta}^{(t+1)}=\arg\min_{\mathbf\Theta_{\boldsymbol\beta}}\big\{\mathcal Q(\mathbf\Theta_{\boldsymbol\beta},\mathbf\Omega_{\boldsymbol\beta}^{(t)},\mathbf\Gamma_{\boldsymbol\beta}^{(t)},\mathbf\Lambda_1^{(t)},\mathbf\Lambda_2^{(t)})\big\},
\label{update1}
\end{align}
where $\mathbf\Omega_{\boldsymbol\beta}^{(t)}$, $\mathbf\Gamma_{\boldsymbol\beta}^{(t)}$, $\mathbf\Lambda_1^{(t)}$, $\mathbf\Lambda_2^{(t)}$ are the estimates at the $t$th iteration. By some algebra calculation, $\mathbf\Theta_{\boldsymbol\beta}^{(t+1)}$ is shown to have a close-form expression $\mathbf\Theta_{\boldsymbol\beta}^{(t+1)}=(-\mathbf Q^{(t)}+\sqrt{\mathbf Q^{(t)}\mathbf Q^{(t)}+8\psi\mathbf I_p})/4\psi$,
where $\mathbf Q^{(t)}=\widehat{\mathbf\Sigma}_{\boldsymbol\beta}+\mathbf\Lambda_{1}^{(t)}+\mathbf\Lambda_{2}^{(t)}-\psi\mathbf\Omega_{\boldsymbol\beta}^{(t)}-\psi\mathbf\Gamma_{\boldsymbol\beta}^{(t)}$. Next, each entry of $\mathbf\Omega_{\boldsymbol\beta}^{(t+1)}$ can be solved separately by:
\begin{align}
\Omega_{\beta_k\beta_s}^{(t+1)}=\arg\min_{\Omega_{\beta_k\beta_s}}\big\{
(\Theta_{\beta_k\beta_s}^{(t+1)}+\Lambda_{ks}/\psi-\Omega_{\beta_k\beta_s})^2+ 2P_{\lambda/\psi}(\Omega_{\beta_k\beta_s})
\big\}.
\label{update2}
\end{align}
If a specific penalty such as MCP is applied, $\Omega_{\beta_k\beta_s}^{(t+1)}$ also has a close-form expression. We discuss the choice of penalty in the next subsection. Next, the update of $\mathbf\Gamma_{\boldsymbol\beta}$ is
\begin{align}
\mathbf\Gamma_{\boldsymbol\beta}^{(t+1)}=\arg\min_{\sigma_\texttt{min}(\mathbf\Gamma_{\boldsymbol\beta})>\delta}\big\{Q(\mathbf\Theta_{\boldsymbol\beta}^{(t+1)},\mathbf\Omega_{\boldsymbol\beta}^{(t+1)},\mathbf\Gamma_{\boldsymbol\beta},\mathbf\Lambda_1^{(t)},\mathbf\Lambda_2^{(t)})\big\},
\label{update3}
\end{align}
and the close-form solution is $\mathbf\Gamma_{\boldsymbol\beta}^{(t+1)}=[\mathbf\Theta^{(t+1)}+\psi\mathbf\Lambda_2^{(t)},\delta]_+$. The updates of $\mathbf\Lambda_1$ and $\mathbf\Lambda_2$ are implemented in the same ways as $\mathbf\Theta_{\boldsymbol\beta}^{(t+1)}$ and others:
\begin{align}
\mathbf\Lambda_1^{(t+1)},\mathbf\Lambda_2^{(t+1)}=\arg\min_{\mathbf\Lambda_1,\mathbf\Lambda_2}\big\{\mathcal Q(\mathbf\Theta_{\boldsymbol\beta}^{(t+1)},\mathbf\Omega_{\boldsymbol\beta}^{(t+1)},\mathbf\Gamma_{\boldsymbol\beta}^{(t+1)},\mathbf\Lambda_1,\mathbf\Lambda_2)\big\},
\label{update4}
\end{align}
which are given by $
\mathbf\Lambda_1^{(t+1)}=\mathbf\Lambda_1^{(t)}+\psi(\mathbf\Theta_{\boldsymbol\beta}^{(t+1)}-\mathbf\Omega_{\boldsymbol\beta}^{(t+1)})$ and $ \mathbf\Lambda_2^{(t+1)}=\mathbf\Lambda_2^{(t)}+\psi(\mathbf\Theta_{\boldsymbol\beta}^{(t+1)}-\mathbf\Gamma_{\boldsymbol\beta}^{(t+1)}).$
The solutions in (\ref{update1}) - (\ref{update4}) are iterated until convergence. The precision matrix estimate $\widehat{\mathbf\Theta}_{\boldsymbol\beta}$ is defined as $\mathbf\Omega_{\boldsymbol\beta}^{(\infty)}$.

\subsection{Implementation Issues}
The penalty function $P_\lambda(\cdot)$ plays a central role in EGG, enabling simultaneous graph estimation and edge selection. Lasso \citep{tibshirani1996regression}, defined as $P^\texttt{lasso}_\lambda(x)=\lambda|x|$, is one of the most common variable selection penalty. However, the lasso introduces bias into the parameter estimation and is inconsistent in variable selection \citep{meinshausen2006high}. To address this issue, nonconvex penalty functions are proposed, which are superior to lasso due to the oracle property \citep{fan2001variable}. One such penalty is the MCP \citep{zhang2010nearly} whose expression is:
\begin{align}
P^\texttt{MCP}_{\lambda,\gamma}(x)=\lambda\int_0^{|x|}\bigg(1-\frac{|x|}{\gamma\lambda}\bigg)_+dx
\end{align}
where $(x)_+=\max(x,0)$ and $\gamma>1$ is an alternative tuning parameter controlling the concavity of $P^\texttt{MCP}_{\lambda,\gamma}(\cdot)$. In particular, the following minimization has a close-form solution:
\begin{align}
\hat\theta=\arg\min_\theta\bigg\{\frac12(x-\theta)^2+P^\texttt{MCP}_{\lambda,\gamma}(\theta)\bigg\}=\begin{cases}
\frac{\gamma}{\gamma-1}\texttt{soft}_\lambda(x),&\text{ if }|x|\leq\lambda\gamma,\\
x,&\text{ if }|x|>\lambda\gamma,
\end{cases}
\end{align}
where $\texttt{soft}_\lambda(x)=\texttt{sign}(x)(|x|-\lambda)_+$ is known as the soft-thresholding operator. Thus, $\Omega_{\beta_k\beta_s}^{(t+1)}$ can be updated with a close-form solution. 

In MCP, two tuning parameters, $\lambda$ and $\gamma$, require appropriate selection. We adhere to the recommendation by \citet{zhang2010nearly} who suggested a universal setting of $\gamma=3$, given that MCP is less sensitive to its choice. As for $\lambda$ that plays a more crucial role, we employ a two-stage selection procedure called stability selection \citep{meinshausen2010stability}. Let $H$ be the number of subsampling times and $\mathcal I^{(h)}$ be a random subsample of ${1,\dots,m}$ of size $[c_sm]$ drawn without replacement. In the first stage, we perform the standard cross-validation scheme to select the optimal $\lambda$, where the cross-validation error is 
\begin{align}
\texttt{CVE}^{(h)}(\lambda)=\texttt{entropy}(\widehat{\mathbf\Sigma}_{\boldsymbol\beta}^{\texttt{test}_{(h)}},\hat{\bm\Theta}_{\boldsymbol\beta}^{\texttt{train}_{(h)}}(\lambda)),
\end{align}
$\hat{\bm\Sigma}_{\boldsymbol\beta}^{\texttt{test}_{(h)}}$ is the genetic covariance estimate with the test dataset, and $\hat{\bm\Theta}_{\boldsymbol\beta}^{\texttt{train}_{(h)}}(\lambda)$ is the genetic precision matrix estimated from train dataset. The optimal $\lambda^\texttt{CV}$ is selected by:
\begin{align}
\lambda^\texttt{CV}=\arg\min_{\lambda\in\mathcal L}\big\{\texttt{CVE}^{(h)}(\lambda)\big\},
\end{align}
where $\mathcal L$ is a set of candidates of $\lambda$. In the second stage, we apply stability selection to reduce the false discovery rates of edge selection. Specifically, we calculate the selection frequency
\begin{align}
\pi_{jk}^{H}=\frac1H\sum_{h=1}^H\texttt{I}\{\hat\Theta_{\beta_k\beta_s}^{(h)}(\lambda^\texttt{CV})\},
\end{align}
and artificially enforce $\hat\Theta_{\beta_k\beta_s}(\lambda^\texttt{CV})=0$ if $\pi_{jk}^{H}<c_t$, where $c_t\in[0.5,1)$ is a threshold. Stability selection is not sensitive to either the subsampling fraction $c_s$ or the threshold $c_t$. In EGG, we set $c_s=0.5$ as this fraction following the suggestion of the authors, and consider $c_t=0.95$ as $1-\pi_{jk}^{H}$ can be used as the empirical P-values of $\hat\Theta_{\beta_k\beta_s}(\lambda^\texttt{CV})$ when $H$ is large enough.

We suggest first estimating the genetic covariance matrix of Z-score $\hat{\boldsymbol z}_j$, denoted as $\widehat{\mathbf\Sigma}_{\hat{\boldsymbol z}}$, and yielding the genetic correlation matrix of the effect sizes by $\widehat{\mathbf R}_{\boldsymbol\beta}=\texttt{cov2cor}(\widehat{\mathbf\Sigma}_{\hat{\boldsymbol z}}-\widehat{\mathbf R}_{\boldsymbol\omega})$. In the literature, the use of Z-scores is a common simplification because each Z-score is a linear transformation of effect size estimates and the variance of the estimation error of the Z-scores is always 1 \citep{bulik2015atlas}. In addition, utilizing a genetic correlation matrix of the Z-scores effectively mitigates the challenges of tuning parameter selections. For example, the choice of $\psi$ depends on the scale of the Hessian matrix of the minimization, and we suggest $\psi\in(0,0.2)$ in EGG as the Hessian matrix of the entropy loss function is $\widehat{\mathbf R}_{\boldsymbol\beta}\otimes\widehat{\mathbf R}_{\boldsymbol\beta}$ whose diagonal entries are all 1. Likewise, $\lambda\in(0,2)$ generally performs well.
\subsection{Convergence Rate and Model Selection Consistency}
In this subsection, we investigate the convergence rates and model selection consistency of EGG.
To facilitate the theoretical derivation, we specify two definitions and six regularity conditions.
\begin{definition}[Sub-Gaussian variable]
A random variable $x$ is sub-Gaussian distributed if there exist a  constant $0<c_0<\infty$ such that for all $t>0$, $\Pr(|x-\texttt{E}(x)|\geq t)\leq 2\exp(-c_0t^2)$. 
\end{definition}
\begin{definition}[Well-conditioned covariance matrix]
A covariance matrix $\bm\Sigma$ is well-conditioned if there is a positive constant $d_0$ such that $0<d_0^{-1}\leq\sigma_\texttt{min}(\bm\Sigma)\leq\sigma_\texttt{max}(\bm\Sigma)\leq d_0<\infty.$
\end{definition}
\begin{condition}[Regularity conditions for random effect model]~
\begin{itemize}
\item[(C1)] For $\boldsymbol G_i=(G_{i1},\dots,G_{im})^\top$, each entry $G_{ij}$ is a sub-Gaussian with $\texttt{E}(G_{ij})$=0 and $\texttt{var}(G_{ij}$)=1. Besides, for all $(i,j)\neq(t,s)$, $G_{ij}$ is independent of $G_{ts}$. Furthermore, there exist a constant $0<\bar{g}<\infty$ such that for all ($i,j$), $\Pr(|G_{ij}|>\bar{g})=0$. 
\item[(C2)] For $\boldsymbol \beta_j=(\beta_{j1},\dots,\beta_{jp})^\top$, $\sqrt m\beta_{js}$ is sub-Gaussian with $\texttt{E}(\sqrt m\beta_{js})=0$ and $\texttt{var}(\sqrt m\beta_{js})\in(0,1)$. Besides, for all $j\neq t$, $\boldsymbol\beta_j$ is independent of $\boldsymbol\beta_t$. Furthermore, $\mathbf\Psi_{\boldsymbol\beta}=m\mathbf\Sigma_{\boldsymbol\beta}=\texttt{cov}(\sqrt m\boldsymbol\beta_j)$ is a well-conditioned covariance matrix.  Finally, $p$ is a fixed number.
\item[(C3)] For $\boldsymbol\epsilon_i=(\epsilon_{i1},\dots,\epsilon_{ip})^\top$, each entry $\epsilon_{ij}$ is a sub-Gaussian with $\texttt{E}(\epsilon_{ij})=0$ and $\texttt{var}(\epsilon_{is})\in(0,1)$. Besides, $\boldsymbol\epsilon_i$ is independent of $\boldsymbol\epsilon_t$  for all $i\neq t$. Furthermore, $\mathbf\Sigma_{\boldsymbol\epsilon}=\texttt{cov}(\boldsymbol\epsilon_i)$ is a well-conditioned covariance matrix.
\item[(C4)] For $\boldsymbol F_i=(F_{i1},\dots,F_{im})^\top$, each entry $F_{ij}$ is a sub-Gaussian with $\texttt{E}(F_{ij})$=0 and $\texttt{var}(F_{ij}$)=1. Besides, for all $(i,j)\neq(t,s)$, $F_{ij}$ is independent of $F_{ts}$. Furthermore, there exist a constant $0<\bar{f}<\infty$ such that for all ($i,j$), $\Pr(|F_{ij}|>\bar{f})=0$. 
\item[(C5)] The genetic variant $\boldsymbol G_i$, the insignificant variant $\boldsymbol F_i$, the genetic effect $\boldsymbol\beta_j$, the noise terms $\boldsymbol\epsilon_i$, are four mutually independent groups. 
\item[(C6)] The penalty $P_\lambda(\cdot)$ is an even function and satisfies $P_\lambda(|x|)$ is increasing and concave in $|x|\in[0,+\infty)$ with $P_\lambda(0)=0$; $P_\lambda(|x|)$ is differentiable in $|x|\in(0,+\infty)$ with $P_\lambda'(0):=P_\lambda'(0+).$ Besides, there exist two constants $0<a_1<a_2<\infty$ such that $P_\lambda'(|x|)\geq a_1\lambda$ for all $|x|\in[0,a_2\lambda]$, and $P_\lambda'(|x|)=0$ for all $|x|\in[a\lambda,+\infty)$, for any $a>a_2$.
\item[(C7)] Let $\mathcal{E}=\{(k,s):\ j\neq k, \Theta_{\beta_j\beta_k}\neq0\}$ where $\mathbf\Theta_{\boldsymbol\beta}=\mathbf\Sigma^{-1}_{\boldsymbol\beta}$, $\Theta_{\beta_j\beta_k}$ is the $(k,s)$th element of $\mathbf\Theta_{\boldsymbol\beta}$, and $\mathcal S=\mathcal E\cup\{(1,1),\dots,(p,p)\}$. Besides,  $\mathbf\Upsilon=\mathbf\Sigma_{\boldsymbol\beta}\otimes\mathbf\Sigma_{\boldsymbol\beta}$ and there is a constant $0< c_\upsilon<\infty$ such that $||\mathbf\Upsilon_{\mathcal S\bar{\mathcal S}}\mathbf\Upsilon_{\mathcal S\mathcal S}^{-1}||_\infty<c_\upsilon$, where the notation $\mathbf\Upsilon_{\mathcal A_1\mathcal A_2}$ refers to a sub-matrix of $\mathbf\Upsilon$ with rows being indices matrix $\mathcal A_1$ and columns in indices matrix $\mathcal A_2$
\end{itemize}
\end{condition}
Conditions (C1)-(C5) ensure that all variables involved in this paper follow a sub-Gaussian distribution. In practice, $G_{ij}$ is standardized from a binomial variable with states 0, 1, and 2. Therefore, it is expected to be a bounded sub-Gaussian variable as long as its minor allele frequency is not rare. Moreover, we assume $\sqrt m\boldsymbol \beta_j$ to be sub-Gaussian with a well-conditioned covariance matrix $\bm\Psi_{\boldsymbol\beta}$, because the cumulative covariance explained by the $m$ IVs $\mathbf\Psi_{\boldsymbol\beta}$ should remain constant while the covariance explained by each IV $\bm\Sigma_{\boldsymbol\beta}\to0$ as $m\to\infty$. Condition (C6) aligns with the standard conditions for variable selection penalties \citep{fan2014strong}. Condition (C7) is known as the irreplaceable condition \citep{ravikumar2011high}, which is crucial for proving the estimation consistency and selection consistency of the EGG. 
\begin{theorem}
Suppose that condition (C1)-(C5) are satisfied. Then there are two constants $0<c_{\Sigma_1}\leq<c_{\Sigma_1}\infty$ such that $\forall t<1$:
\[
\Pr\bigg(\max_{1\leq j\leq k\leq p}|\hat\Sigma_{\beta_j\beta_k}-\Sigma_{\beta_j\beta_k}|\leq t\sqrt{\Sigma_{\beta_j\beta_j}\Sigma_{\beta_k\beta_k}}\bigg)\geq1-\exp\bigg(\log(c_{\Sigma_1}p^2)-n_\texttt{eff}c_{\Sigma_2} t^2\bigg),
\] 
where $n_\texttt{eff}=\min(m,M,n_1,\dots,n_p)$ and $\hat\Sigma_{\beta_j\beta_k}$ is the $(k,s)$th entry of $\widehat{\mathbf\Sigma}^\texttt{Pearson}_{\boldsymbol\beta}$ or $\widehat{\mathbf\Sigma}^\texttt{Spearman}_{\boldsymbol\beta}$. \label{theorem1}
\end{theorem}
Theorem \ref{theorem1} is a pivotal theoretical result. By setting $t=c_{\Sigma_3}\sqrt{(\log n_\texttt{eff})/n_\texttt{eff}}$ with an alternate constant $0<c_{\Sigma_3}<\infty$, we obtain that
\[
\max_{1\leq j\leq k\leq p}\bigg\{\frac{|\hat\Sigma_{\beta_j\beta_k}-\Sigma_{\beta_j\beta_k}|}{\sqrt{\Sigma_{\beta_j\beta_j}\Sigma_{\beta_k\beta_k}}}\bigg\}\leq c_{\Sigma_3}\sqrt{\frac{\log n_\texttt{eff}}{n_\texttt{eff}}}
\]
with a probability exceeding $1-\exp(\log(c_{\Sigma_1}p^2)-c_{\Sigma_2}c_{\Sigma_3}^2\log n_\texttt{eff})$. This is, after accounting for the scale of $\mathbf\Sigma_{\boldsymbol\beta}$, $\widehat{\mathbf\Sigma}^\texttt{Pearson}_{\boldsymbol\beta}$ and $\widehat{\mathbf\Sigma}^\texttt{Spearman}_{\boldsymbol\beta}$ converge at a non-asymptotic rate $O(\sqrt{(\log n_\texttt{eff})/n_\texttt{eff}})$, which approaches 1 as $n_\texttt{eff}$ tends towards infinity. Moreover, this theorem highlights that the effective ``sample size'' for genetic covariance estimation is $n_\texttt{eff}=\min(m,M,n_1,\dots,n_p)$. Given that the sample sizes of GWAS cohorts typically far exceed the number of independent signals reaching genome-wide significance (P-value$<$5E-8), the accuracy of genetic covariance estimation is primarily determined by the number of independent genetic variants used.
\begin{theorem}
Suppose that condition (C1)-(C7) are satisfied. Then there are four constants $0<c_{\lambda_1}\leq c_{\lambda_2}<\infty$ and $0<c_{\Theta_1}\leq c_{\Theta_2}<\infty$ such that $\forall t<1$:
\[
\Pr\bigg(\max_{1\leq j\leq k\leq p}|\hat\Theta_{\beta_j\beta_k}-\Theta_{\beta_j\beta_k}|\leq t\sqrt{\Theta_{\beta_j\beta_j}\Theta_{\beta_k\beta_k}}\bigg)\geq1-\exp\bigg(\log(c_{\Theta_1}p^2)-n_\texttt{eff}c_{\Theta_2} t^2\bigg),\]
if $\lambda=c_{\lambda_1}(\max_{1\leq j \leq p}\Sigma_{\beta_j\beta_j})t$. In addition,
\[
\Pr\bigg(\forall(k,s),\ \texttt{sign}(\hat\Theta_{\beta_j\beta_k})=\texttt{sign}(\Theta_{\beta_j\beta_k})\bigg)\geq1-\exp\bigg(\log(c_{\Theta_1}p^2)-n_\texttt{eff}c_{\Theta_2} t^2\bigg),
\]
if $\min_{1\leq k\leq j\leq p}|\Theta_{jk}|>c_{\lambda_2}(\max_{1\leq k\leq p}\Theta_{\beta_k\beta_k})t$. \label{theorem2}
\end{theorem}
Theorem \ref{theorem2} highlights two key points: the genetic network $\mathbf\Theta_{\boldsymbol\beta}$ can be estimated with the same non-asymptotic convergence rate as $\mathbf\Sigma_{\boldsymbol\beta}$, and the edge set can be consistently recovered with a probability exceeding $1-\exp(\log(c_{\Theta_1}p^2)-n_\texttt{eff}c_{\Theta_2} c_{\Sigma_2} t^2)$. This probability approaches 1 as $n_\texttt{eff}$ tends towards infinity, which assures that both the genetic network estimate $\widehat{\mathbf\Theta}_{\boldsymbol\beta}$ and the edge set estimate $\hat{\mathcal E}=\{(k,s):\ j\neq k, \hat\Theta_{\beta_j\beta_k}\neq0\}$ will be consistent in an asymptotic sense.
\begin{table}[t]
\footnotesize
\centering
\caption{GWAS summary data used in real data analysis.}
\begin{tabular}{ccccc}
&&European&&East Asian\\
Trait&Sample Size  & Source &Sample Size & Source \\
ALB  & 363,228& \citet{sinnott2021genetics} & 217,780 & \citet{nam2022genome}\\
ALT  & 437,267  & \citet{pazoki2021genetic}	& 288,137 & \citet{kim2022contribution}\\
AST  & 437,438 & \citet{pazoki2021genetic}	& 288,137 & \citet{kim2022contribution}\\
BMI  & 669,688   & \citet{loh2018mixed}+MVP	& 236,117 & \citet{nam2022genome}\\
BUN  & 1,201,929 & \citet{stanzick2021discovery}	& 221,053 & \citet{nam2022genome}\\
CAD  & 1,458,128 & \citet{aragam2022discovery}+MVP& 212,453 & \citet{ishigaki2020large}\\
sCr& 363,228& \citet{sinnott2021genetics} & 217,780 & \citet{nam2022genome}\\
GGT  & 437,194 & \citet{pazoki2021genetic}	& 288,137 & \citet{kim2022contribution}\\
HBA1C & 344,182 & Neale's lab	& 288,137 & \citet{kim2022contribution}\\
HDL  & 1,320,016 & \citet{graham2021power} & 288,137 & \citet{kim2022contribution}\\
LDL& 1,320,016 & \citet{graham2021power}	& 288,137 & \citet{kim2022contribution} \\
PLT&542.827& \citet{chen2020trans}&202,552  &\citet{nam2022genome}\\
RBC& 542,827& \citet{chen2020trans} & 207,876 & \citet{nam2022genome}\\
TG & 1,320,016 & \citet{graham2021power}	& 288,137 & \citet{kim2022contribution} \\
SBP& 1,004,643 & \citet{surendran2020discovery} & 217,780 & \citet{nam2022genome} \\
Stroke &1,296,908&\citet{mishra2022stroke} &  256,274 & \citet{mishra2022stroke} \\
T2D& 1,114,458 & \citet{vujkovic2020discovery} & 433,540 & \citet{nam2022genome}\\
UA& 343,836 & Neale's lab & 129,405 &  \citet{kanai2018genetic}\\
WBC&562,243& \citet{chen2020trans} & 208,720   & \citet{nam2022genome}
\end{tabular}
\end{table}
\section{Real Data Analysis}
\subsection{Data Processing}
\begin{sidewaysfigure}[p]
\begin{center}
\includegraphics[width=6.6in]{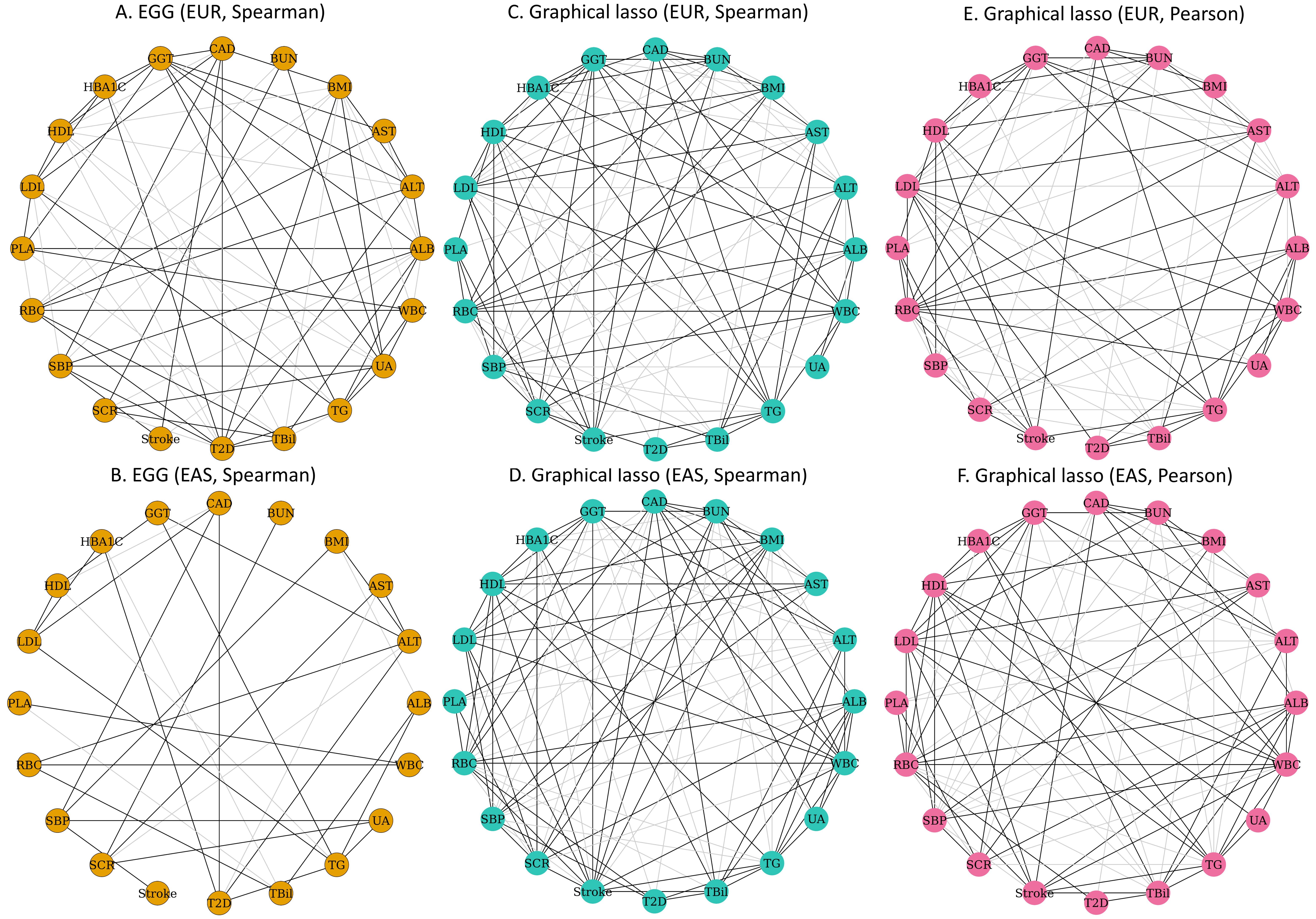}
\end{center}
\caption{ 
Panels A to C illustrate the networks of 20 metabolic and cardiovascular traits for the EUR population, as determined by EGG using Spearman's rho, graphical lasso with Spearman's rho, and graphical lasso with Pearson's r, respectively. Similarly, panels D to F present the equivalent networks for the EAS population. Traits within the same category are color-coded uniformly; for instance, PLT, RBC, and WBC are all depicted in orange. In these networks, a grey edge represents a negative conditional correlation between traits, while a black edge denotes a positive conditional correlation.
\label{fig1}}
\end{sidewaysfigure}
We employ the EGG approach to explore the genetic network of 20 metabolic and cardiovascular traits, including CAD, T2D, stroke, body mass index (BMI), liver function markers such as alanine aminotransferase (ALT), aspartate aminotransferase (AST), 
$\gamma$-glutamyl transferase (GGT), and total bilirubin (TBil), blood sugar metrics glycated hemoglobin (HBA1C), kidney function indicators including serum albumin (ALB), blood urea nitrogen (BUN), serum creatinine (sCr), and uric acid (UA), blood cell counts such as platelet (PLT), red blood cell (RBC), and white blood cell (WBC) counts, serum lipids high-density lipoprotein (HDL), low-density lipoprotein (LDL), and triglycerides (TG), and systolic blood pressure (SBP). Our aim is to identify the robust genetic networks of these traits and to investigate the network difference in EUR and EAS populations. We do not include other highly correlated traits such as blood glucose, diastolic blood pressure, and pulse pressure because inclusion of highly correlated traits results in instability of network estimation. Table 1 provides a summary of the GWAS data utilized in this analysis, representing the largest GWAS sample sizes available to date for the most of traits.

We utilized the LD reference panel from the 1000 genomes project, which includes 2,490 participants and 1.67M variants in common with either HapMap3 \citep{international2010integrating} or the UK Biobank \citep{sudlow2015uk}. We constructed population-specific reference panels using EUR and EAS participants, each with a sample size of around 500. After allele harmonization, we obtained 1.37M variants for the EUR and 1.13M for the EAS. We then focused on variants from the union set and used the non-significant SNPs (i.e., P-value $>0.05$ for all traits) to estimate the correlation matrix $\widehat{\mathbf R}_{\boldsymbol\omega}$. This matrix was subsequently used as the correlation matrix of effect sizes in a joint $\chi^2_{20}$ test. Independent variants were selected from these variants survived after C+T pruning using the $\chi^2_{20}$ test p-values, which yield 5,458 independent variants for EUR and 2,732 independent variants for EAS, with the P-value thresholds 5E-8 and 5E-6, respectively. We adjusted the threshold for the EAS traits downward slightly due to their significantly smaller sample sizes compared to the EUR ones. These independent variants had evidence of association with at least one of the 20 traits and were not in LD. Furthermore, we employed Spearman's rho and MAD to estimate the covariance matrix of the Z-scores and then yielded the genetic correlation matrix of the effect sizes. For other parameters, we refer to the recommendations in Section 3.4.
\subsection{Results}  
Fig \ref{fig1}A presents a genetic Gaussian graph detailing the relationships among 20 metabolic and cardiovascular traits in EUR population. We observed that multiple traits, including BMI, T2D, liver function measure GGT, and lipid levels HDL and LDL, as well as SBP and Stroke, exhibited a direct connection to CAD. Notably, our study suggests the potential direct causal link between liver function and CAD conditional on other traits we included. Besides, the network also revealed a cluster of traits directly linked to T2D, including ALT, BMI, GGT, HBA1C HDL, RBC, TBil, and TG. While certain liver, kidney, and blood cell traits appear to be direct risk factors for T2D, they may not directly influence CAD. As a consequence, the previously observed causal links from RBC and HBA1C to CAD can be explained as the mediation through T2D \citep{wang2022mendelian}. This underlines the critical role of T2D in the development of CAD, potentially mediating the impact of various metabolic traits \citep{ahmad2015mendelian}. Furthermore, it indicated a connection between ischemic stroke and CAD, with SBP emerging as a common risk factor for both conditions. The data analysis also observed a potential protective role of higher ALB levels against ischemic stroke risk. Experimental evidence, including studies using a rat model of transient focal cerebral ischemia, suggests that ALB can prevent stroke \citep{cole1990hypertension}. This protective effect is attributed to hemodilution caused by ALB, leading to decreased hematocrit, reduced infarct volume, and less cerebral edema. 

Interestingly, We observed a negative link between BMI and SBP, which seems inconsistent with what we understand. This discrepancy can arise from the inclusion of BMI as a covariate in the blood pressure GWAS. Ideally, the genetic correlation between SBP and BMI should be zero when BMI is adjusted in SBP GWAS. However, our EGG analysis considered multiple traits simultaneously. The connection between BMI and SBP can be viewed as an association after adjusting for the rest variables that are correlated with BMI. Therefore, a negative connection can happen.  We suggest that it is more reasonable to construct a network based on the GWAS summary statistics that are calculated without adjusting for any traits that are also included in the network analysis. For example, we should use SBP GWAS summary statistics without adjusting for BMI. However, in SBP GWAS, it is common to perform GWAS by adjusting for BMI.  

Fig \ref{fig1}B presents the network in the EAS population using the same set of traits as the EUR population. Consistent with the European network, CAD was directly connected with T2D, HDL, LDL, and SBP. As for T2D, only ALT, AST, HBA1C, and TG showed direct associations. CAD and stroke were associated through their common risk factor SBP, and the protective effect of ALB for stroke was not observed in EAS. We attribute part of the difference between EUR and EAS networks to the smaller sample sizes and less statistical power in the EAS GWAS than the EUR GWAS. Theorem \ref{theorem2} suggests that the number of independent variants $m$ governs the estimation error of network estimate. Increasing EAS GWAS sample sizes will help uncover more causal variants, improving network analysis. Interestingly, we did not observe a negative connection between BMI and SBP, which can be attributed that SBP GWAS in EAS did not adjust for BMI. 
\subsection{Comparison between EGG and Graphical Lasso}
\begin{figure}[t]
\begin{center}
\includegraphics[width=4.7in]{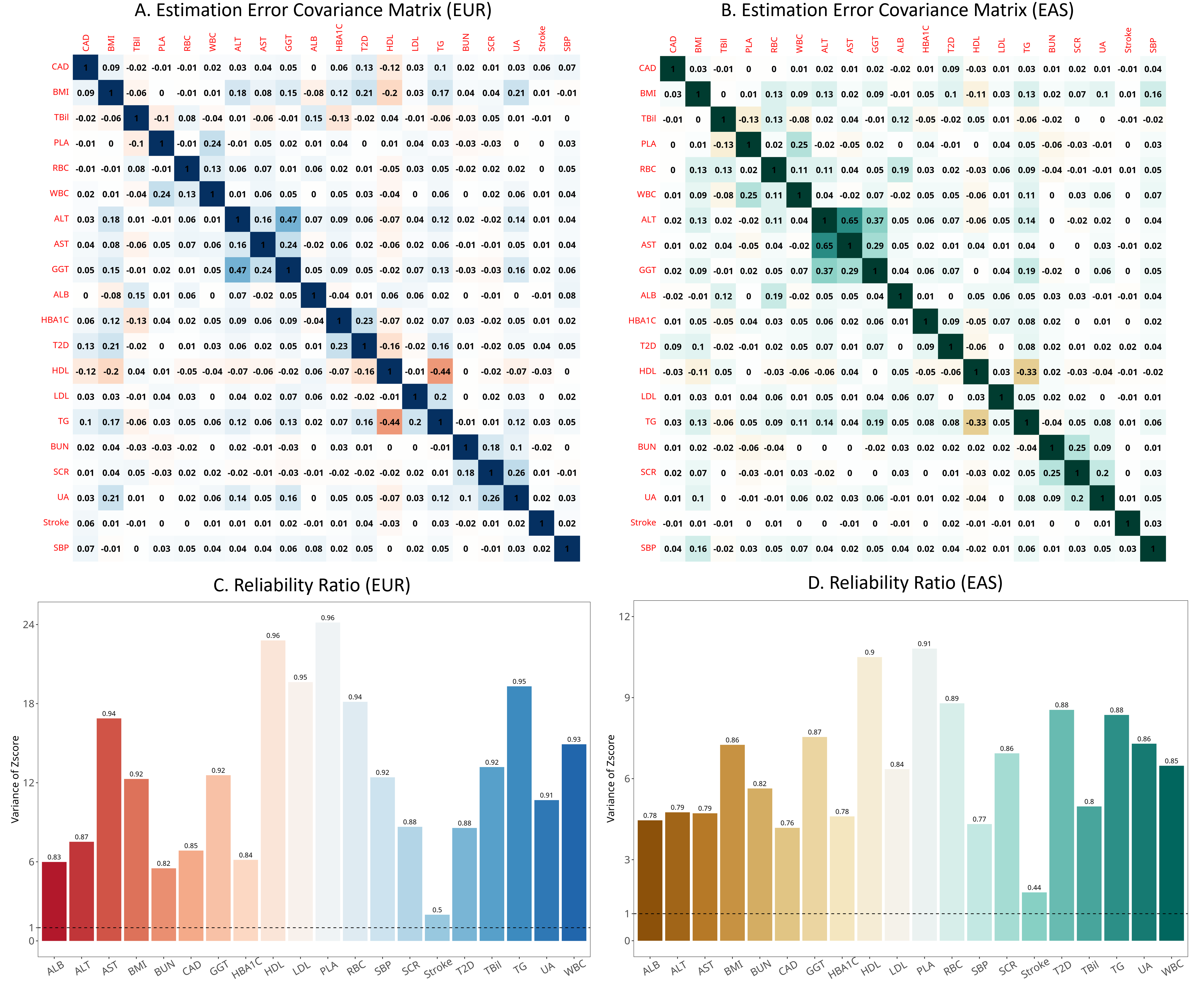}
\end{center}
\caption{ 
Panels A and B show the covariance matrix estimates of estimation errors for EUR and EAS. Panels C and D demonstrate the reliability ratio of each trait that indicates the proportion of variability in the GWAS effect estimates attributable to the true genetic effect. 
\label{fig2}}
\end{figure}   
To show the practical significance of EGG, we compared it with the graphical lasso \citep{friedman2008sparse} with $\mathbf\Sigma_{\boldsymbol\beta}^\texttt{Spearman}$ and $\mathbf\Sigma_{\boldsymbol\beta}^\texttt{Pearson}$ as inputs. To ensure fairness, both EGG and graphical lasso were subjected to stability selection to choose the optimal lambda and employed subsampling to reduce type I errors. Panels C - F in Fig \ref{fig2} present the results. The analysis revealed that the graphical lasso in both EUR and EAS leads to denser networks than EGG and the corresponding interpretation is more challenged, including multiple biologically implausible edges, such as a positive risk correlation between HDL and stroke, and a negative risk correlation between LDL and CAD. The connection between T2D and CAD disappears in the networks estimated by graphical lasso.

We consider two main factors contributing to these discrepancies. First, common tuning parameter selections like CV, which aim to find the optimal parameters for the best prediction, will asymptotically lead to inconsistent model selection \citep{meinshausen2006high}. In contrast, non-convex penalties such as the MCP have been theoretically proved to ensure consistent model selection \citep{fan2001variable}. As EGG employs MCP while graphical lasso uses lasso, the network generated by EGG tends to be more precise. Second, the graphical lasso did not adequately address the estimation error bias. This issue is evident in Fig \ref{fig2}A-\ref{fig2}B. Since the primary samples for EUR and EAS were from the UK Biobank and Biobank Japan, respectively, there are considerable sample overlaps which consequently leads to significant correlations in estimation errors. Fig \ref{fig2}C-\ref{fig2}D, which demonstrate the reliability ratio of the GWAS summary data, further support this point. This ratio, calculated as $\sum_{j=1}^m(\hat z_{jk}^2-1)/\sum_{j=1}^m\hat z_{jk}^2$, reflects the proportion of variance due to genetic effects versus estimation errors \citep{yi2017statistical}. For EUR and EAS, our results indicate that genetic effects may contribute to only about 90\% and 80\% of the total variance of the Z-scores used to explore the networks, respectively, highlighting the significance of estimation errors. Therefore, EGG is likely to yield more interpretable genetic network estimates than graphical lasso by effectively accounting for the estimation error bias.
\section{Simulation}
\subsection{Simulation Settings} 
\begin{figure}[t]
\begin{center}
\includegraphics[width=6in]{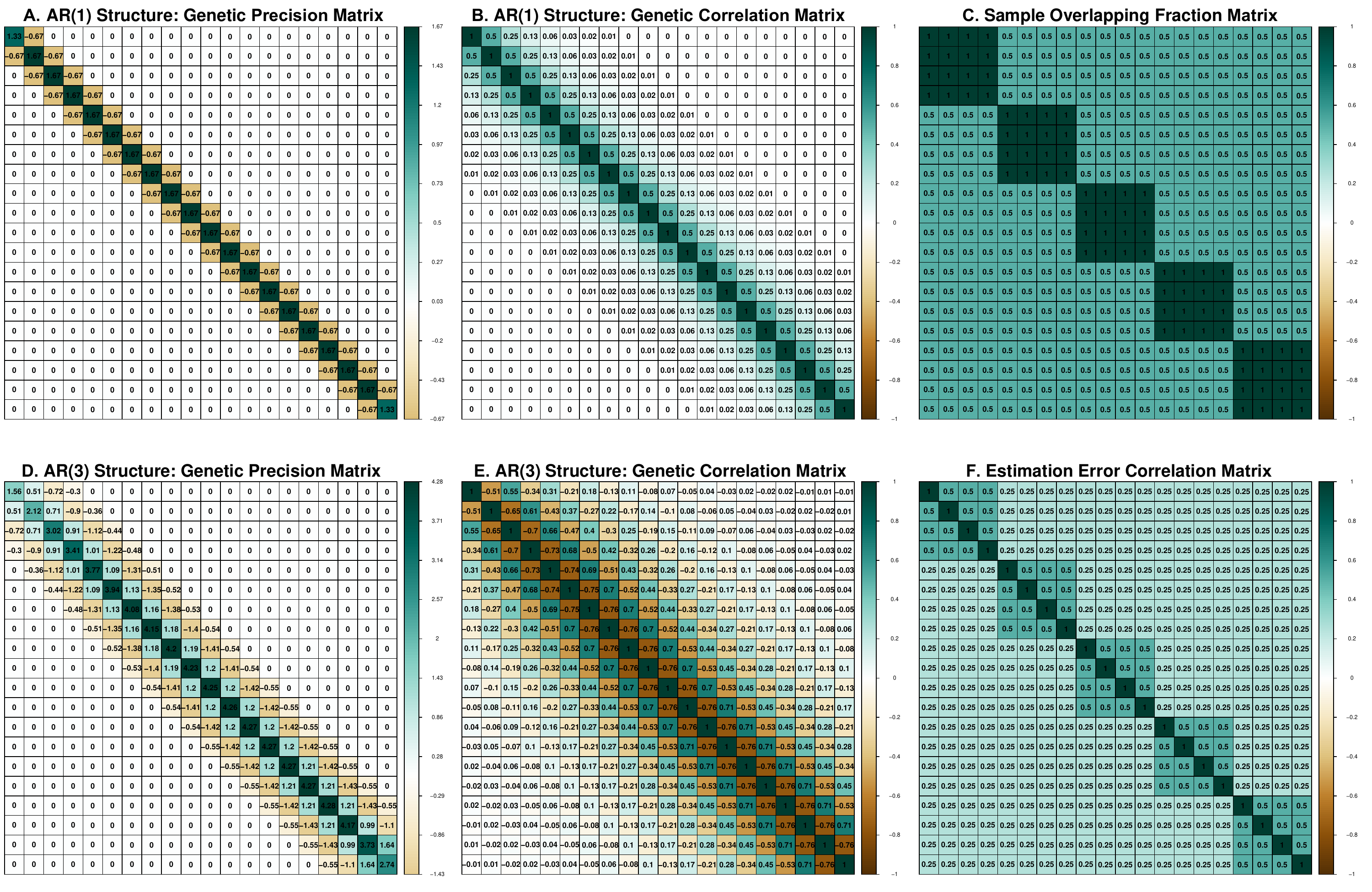}
\end{center}
\caption{Panel A displays the true genetic precision matrix with an AR(1) structure. Panel B presents the corresponding genetic correlation matrix with the AR(1) structure. Panel C illustrates the sample overlapping fractions among 20 traits. Panels D and E exhibit the counterparts of Panels A and B but follow an AR(3) structure. Panel F demonstrates the correlation matrix of estimation errors.
\label{figsimu}}
\end{figure}  
We consider two structures of the genetic precision matrix $\mathbf\Theta_{\boldsymbol\beta}$: AR(1) structure and AR(3) structure, and the sample overlap matrix $\mathbf R_\texttt{overlap}$ is given by $\mathbf N^\texttt{overlap}$ is of a kronecker structure. The dimension of $\widehat{\mathbf\Theta}_{\boldsymbol\beta}$ is $p=20$ which is consistent with the real data. The direct estimate of genetic covariance matrix $\texttt{cov}(\hat{\boldsymbol\beta}_j)=\texttt{cov}(\boldsymbol\beta_j)+\texttt{cov}(\boldsymbol\omega_j)$,
where the covariance of estimation error is approximately $\texttt{cov}(\omega_{jk},\omega_{js})=\texttt{cov}(X_{ik},X_{is})\times n_{jk}/\surd (n_jn_k)$, where $n_{jk}/\surd(n_jn_k)$ is defined as overlapping fraction. Here, we consider $\texttt{cov}(X_{ik},X_{is})=0.5$ for all traits. Fig \ref{figsimu} shows the structures of involved matrices. 

In our study, we examined two scenarios: one with no pleiotropy and another with 10\% pleiotropy. For the latter, we followed existing literature to introduce a mean shift five times the effect size for a randomly selected 10\% of the independent variants \citep{avella2018robust}. We employed both Pearson's r method and Spearman's rho method to estimate $\widehat{\mathbf\Sigma}_{\boldsymbol\beta}$, where Pearson's method would perform poorly in the presence of pleiotropy. We considered three different sample sizes $n=$50K, 200K, 800K for all traits and three different numbers of independent variants $m=500$, 1000, 2000, which collectively explained 20\% of the heritability in each trait. We use the entropy loss function $\texttt{entropy}(\mathbf\Sigma_{\boldsymbol\beta},\widehat{\mathbf\Theta}_{\boldsymbol\beta})$ and the quadratic loss function 
\[
\texttt{Quadratic}(\mathbf\Sigma_{\boldsymbol\beta},\widehat{\mathbf\Theta}_{\boldsymbol\beta})=\texttt{tr}\{(\mathbf\Sigma_{\boldsymbol\beta}\widehat{\mathbf\Theta}_{\boldsymbol\beta}-\mathbf I)^\top(\mathbf\Sigma_{\boldsymbol\beta}\widehat{\mathbf\Theta}_{\boldsymbol\beta}-\mathbf I)\}\]
to assess the estimation error, where $\mathbf\Sigma_{\boldsymbol\beta}$ is the true genetic correlation matrix and $\widehat{\mathbf\Theta}_{\boldsymbol\beta}$ is the related genetic precision matrix estimate. We additionally considered the following ratios:
\begin{align}
t_1=\frac{\#\{(k,s),\ \hat\Theta_{\beta_k\beta_s}\neq0,\ (k,s)\not\in\mathcal{E}\}}{p^2},\quad t_2=\frac{\#\{(k,s),\ \hat\Theta_{\beta_k\beta_s}=0,\ (k,s)\in\mathcal{E}\}}{p^2},
\end{align}
where $t_1$ represents the proportion of false positive edges to the total number of elements in the matrix, while $t_2$ represents the proportion of false negative edges to the total number of elements in the matrix. These can serve as measures for the Type-I and Type-II error rates, respectively. We compare EGG with the graphical lasso (Glasso) \citep{friedman2008sparse}, the constrained $\ell_1$-minimization for inverse matrix estimation (CLIME) \citep{cai2011constrained}, and penalized D-trace estimation (DLasso) \citep{zhang2014sparse}. We employed stability selection of the tuning parameters for EGG, contrasting with the Bayesian information criterion \citep{schwarz1978estimating} used in alternative approaches. The number of replications was 1000.
\subsection{Results}
\begin{figure}[t]
\begin{center}
\includegraphics[width=5.15in]{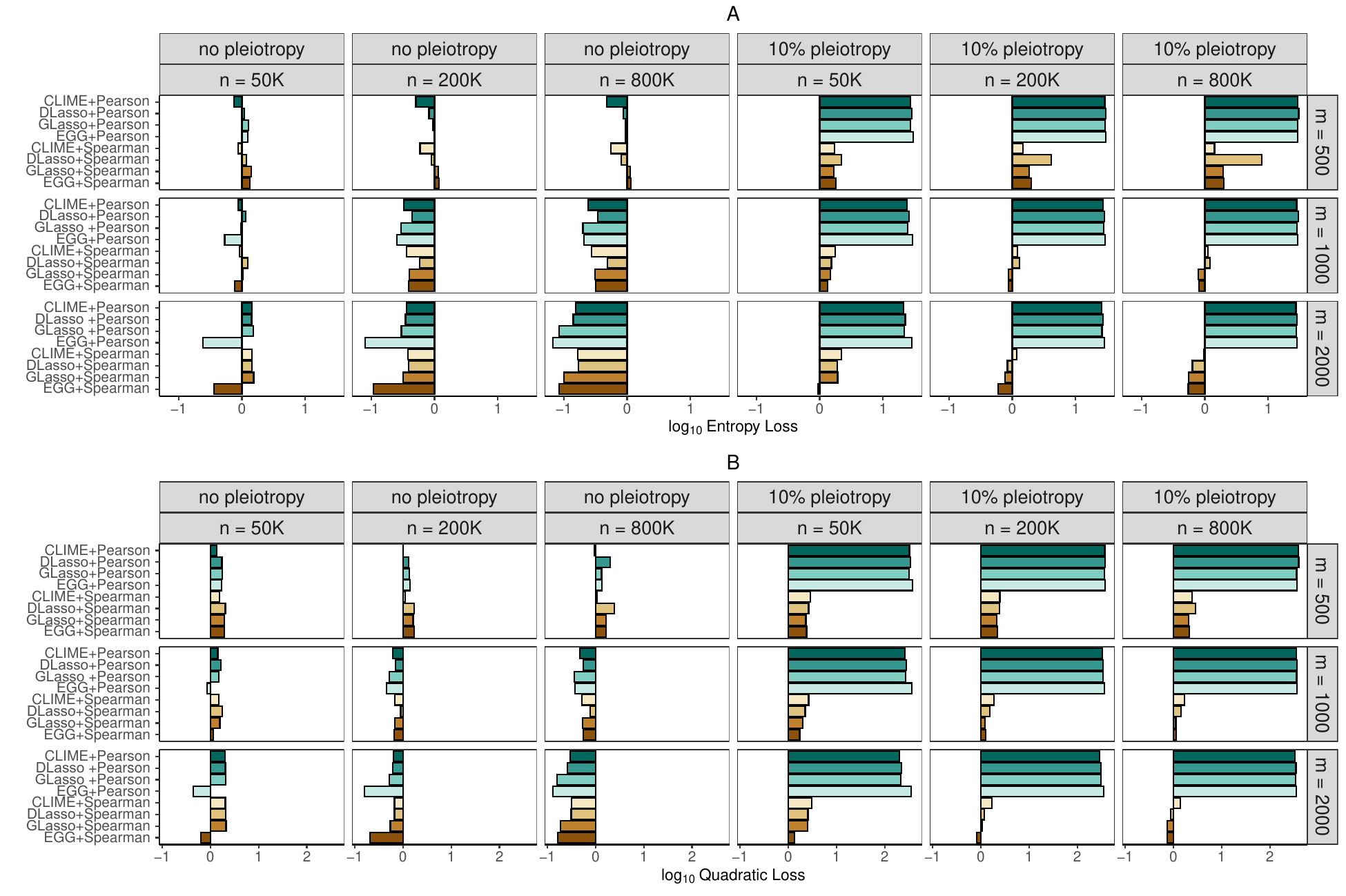}
\end{center}
\caption{Panel A displays the entropy loss of the involved genetic precision matrix estimates in logarithmic scale of the AR(3) structured precision matrix. Panel B shows the corresponding quadratic loss in logarithmic scale.
\label{fig3}}
\end{figure}   
\begin{figure}[t]
\begin{center}
\includegraphics[width=5.15in]{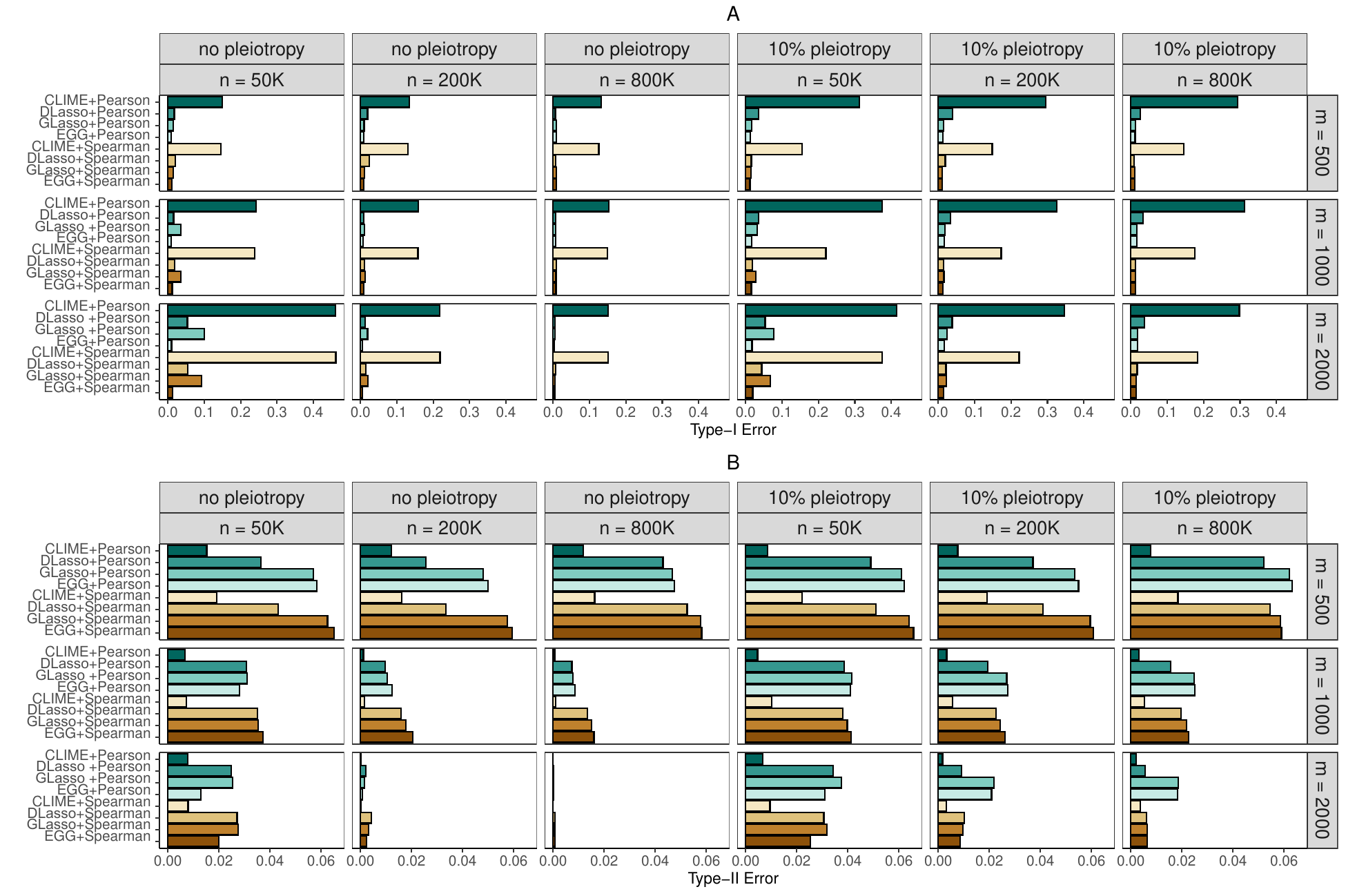}
\end{center}
\caption{Panel A and B respectively displays the measures of Type-I and Type-II error rates $t_1$ and $t_2$ of the AR(3) structured precision matrix.
\label{fig4}}
\end{figure} 

Fig \ref{fig3} displays the bar plots of the two criteria for estimation errors: entropy loss and quadratic loss. We apply a base-10 logarithmic transformation to the estimation errors to make results more discernible. Fig \ref{fig4} illustrates the bar plots for the criteria of type-I and type-II errors, denoted as $t_1$ and $t_2$. The closer the value is to 0, the more capable the network method is of replicating the true graph structure. 

Fig \ref{fig3} indicates that increasing $m$ has a more significant impact on reducing the network estimation error than increasing $n$. This aligns well with our theoretical expectations that the estimation error is determined by $\surd(\log n_\texttt{eff}/n_\texttt{eff})$ where $n_\texttt{eff}=\min(m,M,n_1,\dots,n_k)$. For the genetic architecture that follows a random effect model, which assumes that the genetic effects of variants approximate a normal distribution across the whole genome, increasing the GWAS sample size primarily helps us identify more causal variants, thereby improving the precision of the genetic precision matrix. In practice, independent variants passing the C+T selection can be regarded as causal variants. This means, we may need to increase the sample size such that we can identify more causal variants, making the genetic network estimated by EGG more precise.

When the model does not have pleiotropy, the EGG estimate based on $\widehat{\mathbf\Sigma}_{\boldsymbol\beta}^\texttt{Pearson}$ yields the smallest estimation error, slightly outperforming the EGG estimate based on $\widehat{\mathbf\Sigma}_{\boldsymbol\beta}^\texttt{Spearman}$. The reason is that the EGG accounts for bias caused by estimation errors, while other methods such as CLIME, GLasso, and DLasso do not. In the presence of pleiotropy, non-robust Pearson's r-based estimates exhibited significant estimation errors, likely due to the higher sensitivity of covariance to outliers. Even in such cases, EGG still performed the best in our simulations since it corrects for biases caused by estimation errors and outliers.

The EGG consistently had the lowest Type-I error rates for identifying edges, which can be attributed to its use of stability selection. For Type-II errors, EGG performed similarly to GLasso and DLasso, suggesting that stability selection does not decrease statistical power when compared with the BIC criterion. CLIME's lower Type-II error rate can be attributed to its higher false discovery rate. Overall, EGG maintained low levels of both types of errors, while CLIME had a higher rate of false positives.

\begin{figure}[t]
\begin{center}
\includegraphics[width=5.15in]{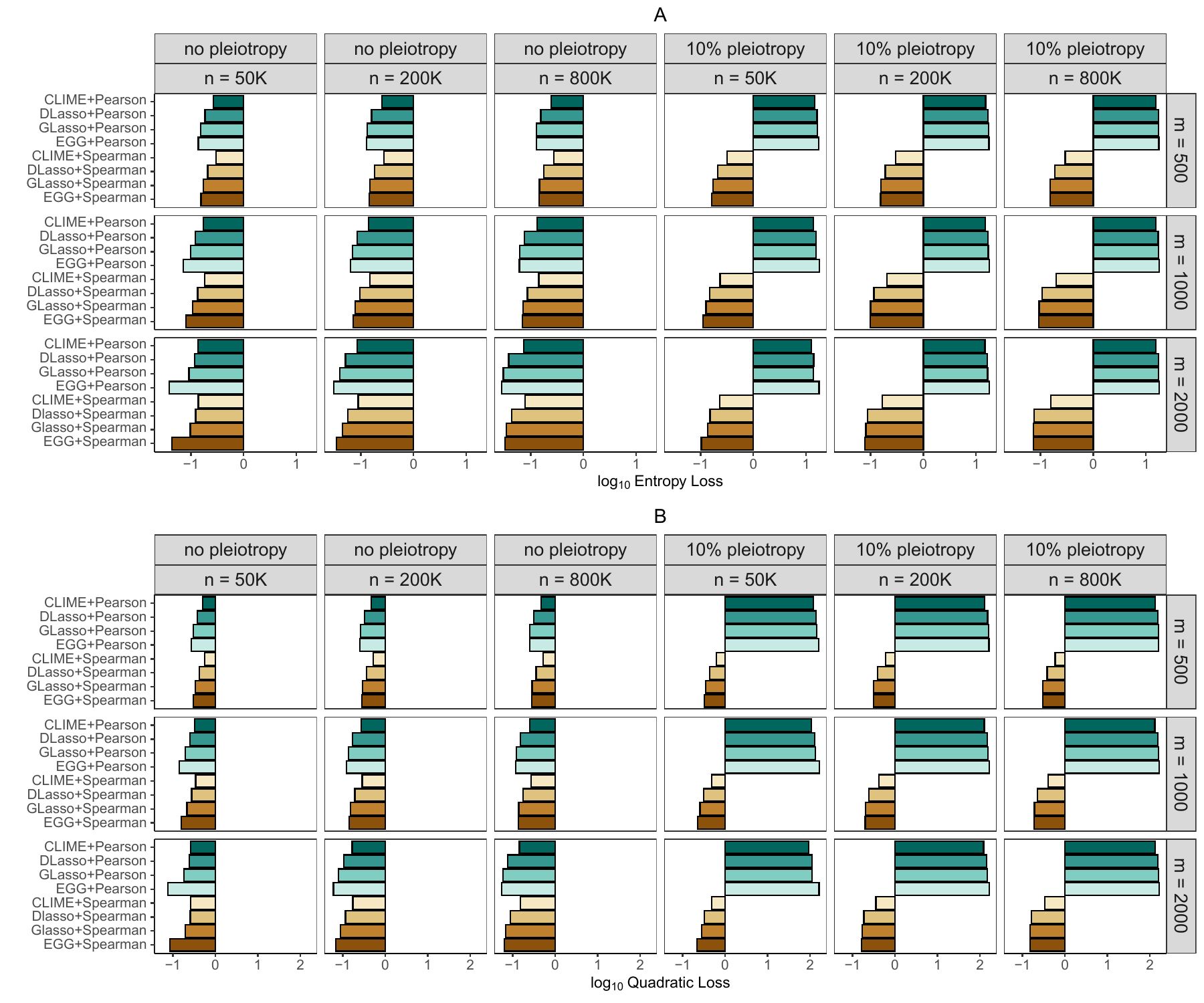}
\end{center}
\caption{Panel A displays the entropy loss of the involved genetic precision matrix estimates in logarithmic scale of the AR(1) structured precision matrix. Panel B shows the corresponding quadratic loss in logarithmic scale.
\label{fig5}}
\end{figure}

\begin{figure}[t]
\begin{center}
\includegraphics[width=5.15in]{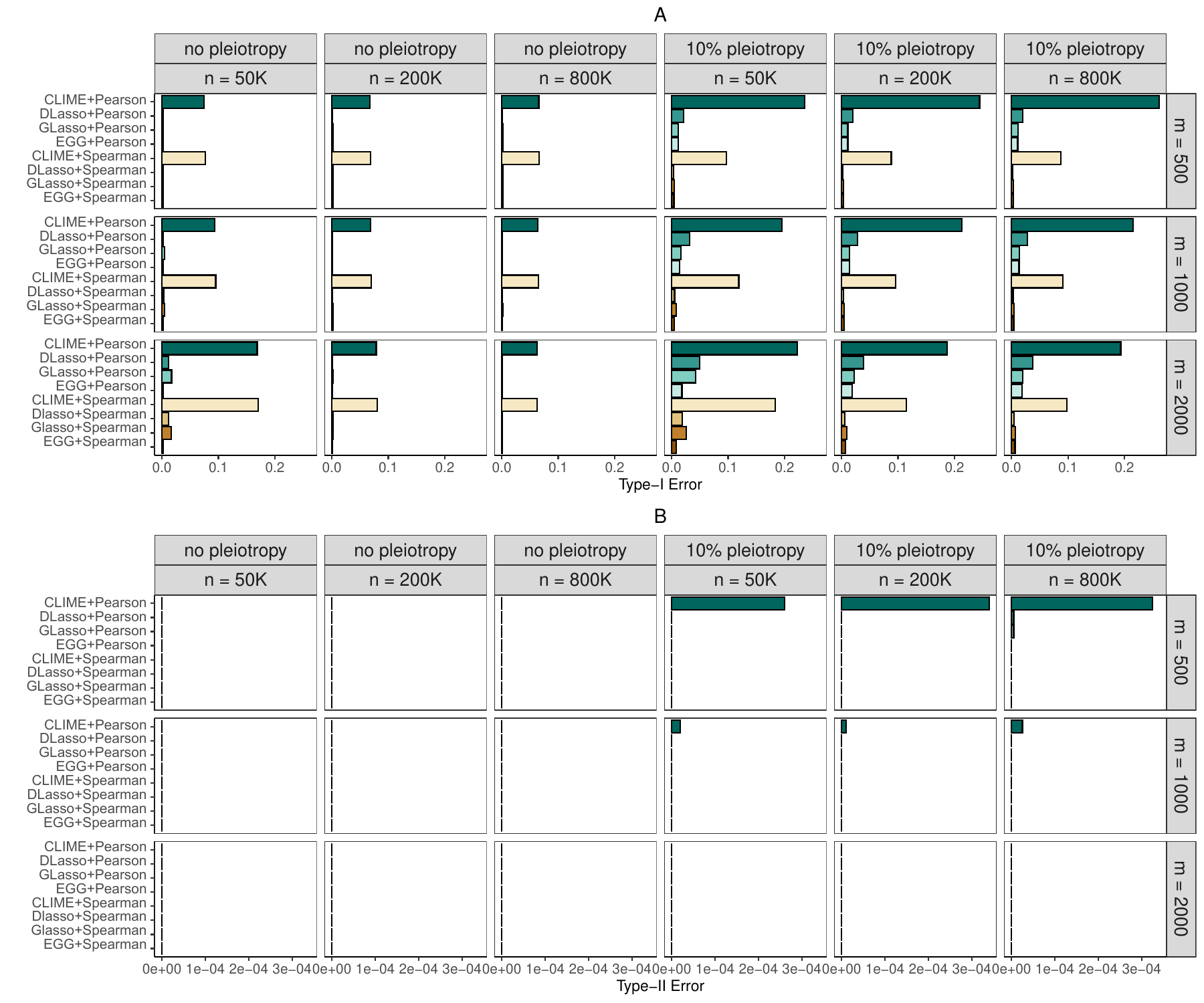}
\end{center}
\caption{Panel A and B respectively displays the measures of Type-I and Type-II error rates $t_1$ and $t_2$ of the AR(1) structured precision matrix.
\label{fig6}}
\end{figure}

Fig \ref{fig5} and Fig \ref{fig6} demonstrate the counterparts of Fig \ref{fig3} and Fig \ref{fig4} in the main body of our paper. In summary, their results are akin to those of the Structure AR(3) model: that is, MGG consistently performs the best under various scenarios, although the difference with existing methods isn't substantial when there's no pleiotropy. Under the AR(1) structure, the measures $t_1$ and $t_2$ for Type-I and Type-II error rates are both notably smaller. This can be attributed to the simpler structure of AR(1), which lacks particularly small elements, making it easier to identify true values compared to the AR(3) structure.
\section{Discussion} 
In this paper, we present the EGG, a novel method that estimates the genetic network of multiple phenotypes by using publicly available summary statistics from GWAS. EGG is robust to the standard biases of MR including weak instruments, horizontal pleiotropy, and sample overlap by employing bias-correction for GWAS estimation error and robust genetic covariance estimation. This is the key component to make EGG superior to traditional network modeling methods. In our study, we examined Gaussian networks for 20 cardiovascular and metabolic traits, including CAD and T2D, across both EUR and EAS populations. Our findings reveal that T2D serves as a direct risk factor for CAD in both populations, underscoring the potential for synergistic treatment strategies for CAD and T2D. For both EUR and EAS, HDL acts as a direct protective factor against CAD, aligning with recent pharmaceutical trial outcomes \citep{hps32017effects}. In addition, we observe that for most metabolic traits, their influences on CAD are indirect through the mediation of blood pressure, lipids levels, and T2D. Through this real data analysis, we demonstrated that EGG represents a potentially significant advancement in both biostatistics and epidemiology, offering new insights into complex causal networks involving multiple disease traits and the related risk factors.

In biology and genetics, there are multiple definitions of phenotype network, potentially causing confusion. According to the review \citep{wang2014review}, three statistical networks are distinguished: marginal correlation network, partial correlation network, and Bayesian network. Marginal correlation network is undirected, based on marginal correlations between phenotypes to explore ``guilt by association'', such as WGCNA used in gene co-expression analysis \citep{langfelder2008wgcna}. Partial correlation network, also undirected, derives from the partial correlation coefficient or the precision matrix, assessing phenotype independence conditional on other phenotypes. Gaussian network belongs to the class of partial correlation networks where the phenotypes are supposedly multivariate-normal distributed \citep{lauritzen1996graphical}. Indeed, GGM may be currently the most favored network approach, owing to its well-studied statistical properties and the availability of powerful tools like graphical lasso \citep{friedman2008sparse} for estimation with individual-level data. The EGG method, as proposed, is novel in estimating Gaussian network from GWAS summary data, uniquely addressing estimation error bias and pleiotropy bias inherent in such data.

Bayesian network is a directed network statistically defined by a structural equation model. Currently, Bayesian network is the major technique used to represent causal diagram among phenotypes under some regularity conditions \citep{pearl2009causality}. However, while Bayesian network holds the potential to uncover complex phenotypic relationships more effectively than partial correlation network, it suffers from substantial optimization challenges \citep{zheng2018dags}, and the complexity in validating the related regularity conditions can significantly reduce the interpretability of Bayesian network estimate \citep{lu2023introducing}. Network deconvolution is a fourth method for network analysis, sharing the same goal of elucidating the directed network of multiple phenotypes as Bayesian network \citep{lin2023combining}. However, it is not categorized among the three mainstream classes of network methods, mainly because its statistical properties have not been as thoroughly researched as those of the aforementioned methods \citep{Pachter2014NetworkNonsense}. Thus, directly comparing network deconvolution with GGM or Bayesian network might be premature, despite its growing popularity. Currently, no established method exists for estimating a Bayesian network of multiple phenotypes using GWAS summary data. Hence, investigating and adapting the EGG approach for this purpose represents a promising and innovative research direction. Additionally, delving into the statistical properties of network deconvolution is crucial, as it will enhance the clarity and biological interpretation of the network deconvolution estimate.

\section*{Proofs}
\subsection{Lemmas}
In this subsection, we specify some lemmas that can facilitate the proofs, most of which can be found in the existing papers. We first discuss the equivalent characterizations of sub-Gaussian (subGau) and sub-exponential (subExp) variables.
\begin{lemma}[Equivalent characterizations of sub-Guassian variables] Given any random variable $X$, the following properties are equivalent:
\begin{itemize}
\item[(I)] there is a constant $K_1\geq0$ such that
\[
\Pr(|X|\geq t)\leq 2\exp(-t^2/K_1^2),\quad\text{for all }t\geq0,
\]
\item[(II)] the moments of $X$ satisfy 
\[
||X||_{L_p}=(\text{E}(|X|^p))^{\frac1p}\leq K_2\surd p,\quad\text{for all }p\geq1,
\]
\item[(III)] the moment generating function (MGF) of $X^2$ satisfies:
\[
\text{E}\{\exp(\lambda^2X^2)\}\leq\exp(K_3^2\lambda^2),\quad\text{for all $\lambda$ staisfying }|\lambda|\leq K_3^{-1},
\]
\item[(IV)] the MGF of $X^2$ is bounded at some point, namely
\[
\text{E}\{\exp(X^2/K_4^2)\}\leq2,
\]
\item[(V)] if E$(X)=0$, the MGF of $X$ satisfies
\[
\text{E}\{\exp(\lambda X)\}\leq\exp(K_5^2\lambda^2),\quad\text{for all }\lambda\in\mathbb{R},
\]
\end{itemize}
where $K_1,\dots,K_5$ are certain strictly positive constants.\label{lemma1}
\end{lemma}
This lemma summarizes some well-known properties of sub-Guassian and can be found in \citet[Proposition 2.5.2]{vershynin2018high}. \textbf{Specifically, we call $K_1$ is the subGau parameter of $X$.}
\begin{lemma}[Equivalent characterizations of sub-exponential variables] Given any random variable $X$, the following properties are equivalent:
\begin{itemize}
\item[(I)] there is a constant $K_1\geq0$ such that
\[
\Pr(|X|\geq t)\leq 2\exp(-t/K_1),\quad\text{for all }t\geq0,
\]
\item[(II)] the moments of $X$ satisfy 
\[
||X||_{L_p}=(\text{E}(|X|^p))^{\frac1p}\leq K_2 p,\quad\text{for all }p\geq1,
\]
\item[(III)] the moment generating function (MGF) of $|X|$ satisfies:
\[
\text{E}\{\exp(\lambda|X|)\}\leq\exp(K_3\lambda),\quad\text{for all $\lambda$ staisfying }0\leq\lambda\leq K_3^{-1},
\]
\item[(IV)] the MGF of $|X|$ is bounded at some point, namely
\[
\text{E}\{\exp(|X|/K_4)\}\leq2,
\]
\item[(V)] if E$(X)=0$, the MGF of $X$ satisfies
\[
\text{E}\{\exp(\lambda X)\}\leq\exp(K_5^2\lambda^2),\quad\text{for all }\lambda\leq K_5^{-1},
\]
\end{itemize}
where $K_1,\dots,K_5$ are certain strictly positive constants.\label{lemma2}
\end{lemma}
This lemma summarizes some well-known properties of sub-exponential and can be found in \citet[Proposition 2.7.1]{vershynin2018high}. \textbf{Specifically, we call $K_1$ is the subExp parameter of $X$.}

\begin{lemma}[Product of sub-Gaussian variable is sub-exponential] Suppose that $X,Z$ are two sub-Gaussian variable, then $Y=XZ$ is a sub-exponential variable. Besides, if $X$ is a bounded sub-Gaussian variable, then then $Y=XZ$ is a sub-Gaussian variable.\label{lemma3}
\end{lemma}
The first claim of this lemma is provided by \citet[Proposition 2.7.7]{vershynin2018high}. The second claim is proved as follows:
\begin{align}
\Pr(Y\geq t)\leq\Pr)(X\geq \bar{X})+\Pr(Z\geq (t/\bar{X}))=\Pr(Z\geq (t/\bar{X}))\leq\exp(-t^2/(\bar{X}^2K_1^2)),
\end{align}
where $\bar{X}$ is the bound of $|X|$ and $K_1$ is the subGau parameter of $Z$. It should be pointed out that if $K_1$ is the subGau/subExp parameter of random variable $X$, then for any constant $K_1'>K1$, $K_1'$ is also the subGau/subExp parameter of $X$. The minimum subGau and subExp parameter for a random variable is defined by
\begin{align}
||X||_{\psi_2}=\inf\{t>0:\ \texttt{E}(X^2/t^2)\leq2\}\\
||X||_{\psi_1}=\inf\{t>0:\ \texttt{E}(|X|/t)\leq2\}.
\end{align}
See  \citet[Chapter 2]{vershynin2018high} for more details.

\subsection{Preliminary properties of random effect model}
We deduce some basic properties of the random effect model. Recall that a multivariate variable $\boldsymbol X_i=(X_{i1},\dots,X_{ip})^\top$ with 
$$
X_{ij}=\eta_{ij}+\epsilon_{ij},
$$
where $\eta_{ij}$ is a part of $X_{ij}$ determined by its genotype and $\epsilon_{ij}$ is a modifiable effect orthogonal to the genetic effect $\eta_{ij}$. The random effect model assumes 
$$
\eta_{ik}=\sum_{j=1}^mG_{ij}\beta_{jk}=\boldsymbol G_i^\top\boldsymbol\beta_k,
$$
where $\boldsymbol G_i=(G_{i1},\dots,G_{im})^\top$ are $m$ genetic variants, $\boldsymbol\beta_k=(\beta_{j1},\dots,\beta_{jp})^\top$ are $m$ genetic effects on the $k$th trait. Besides, $\boldsymbol G_i$ and $\boldsymbol\beta_j$ are considered mutually independent and follow the sub-Gaussian distributions defined in conditions (C1) and (C2). Next,
$$
\texttt{cov}(\boldsymbol X_i)=\mathbf\Sigma_{X},
$$
whose diagonal elements are all 1. Finally, we set
$$
\Pr(|\beta_{jk}|\geq t)=\Pr(|\sqrt m\beta_{jk}|\geq \sqrt mt)\leq2\exp(-mt^2/\kappa_\beta^2),
$$
$$
\Pr(|\epsilon_{ik}|\geq t)\leq2\exp(-t^2/\kappa_\epsilon^2)
$$
where $\kappa_\beta$ is the subGau parameter for $\sqrt m\beta_{jk}$, $1\leq j\leq m$, $1\leq k\leq p$, and $\kappa_\epsilon$ is the subGau parameter for $\epsilon_{ik}$, $1\leq i\leq n_k$, $1\leq k\leq p$. The bounds of $|G_{ij}|$ and $|F_{ij}|$ are $\bar{g}$ and $\bar{f}$, respectively.

We first show
\begin{align}
\Pr(|G_{ij}\beta_{jk}|\geq t)\leq\Pr(|G_{ij}|\geq \bar{g})+\Pr\bigg(|\beta_{jk}|\geq \frac{t}{\bar{g}}\bigg)\leq0+2\exp\bigg(-\frac{mt^2}{\bar{g}^2\kappa_\beta^2}\bigg),
\end{align}
which indicate that $\sqrt mG_{ij}\beta_{jk}$ is a subGau variable with a subGau parameter $\bar{g}\kappa_\beta$. Next,
\begin{align}
&\Pr(|\eta_{ij}|\geq t)=\Pr\bigg(\bigg|\sum_{j=1}^mg_{ij}\beta_{jk}\bigg|\geq t\bigg)\leq2\exp\bigg(-\frac{t^2}{\bar{g}^2\kappa_\beta^2}\bigg),\\
&\Pr(|\eta_{ij}-G_{ij}\beta_{jk}|\geq t)=\Pr\bigg(\bigg|\sum_{z\neq j}^mg_{iz}\beta_{zk}\bigg|\geq t\bigg)\leq2\exp\bigg(-\frac{m}{m-1}\frac{t^2}{\bar{g}^2\kappa_\beta^2}\bigg)\leq2\exp\bigg(-\frac{t^2}{\bar{g}^2\kappa_\beta^2}\bigg),
\end{align}
which indicates $\eta_{ij}$ and $\eta_{ij}-G_{ij}\beta_{jk}$ are two subGau variables with subGau parameter $\bar{g}\kappa_\beta$.  Next, 
\begin{align}
&\Pr(|X_{ik}|\geq t)=\Pr(|\eta_{ik}+\epsilon_{ik}|\geq t)\leq2\exp\bigg(-\frac{t^2}{C_x\bar{g}^2\kappa_\beta^2+C_x\kappa_\epsilon^2}\bigg),\label{subx}\\
&\Pr(|X_{ik}-G_{ij}\beta_{jk}|\geq t)=\Pr(|\eta_{ik}-G_{ij}\beta_{jk}+\epsilon_{ik}|\geq t)\leq2\exp\bigg(-\frac{t^2}{C_x\bar{g}^2\kappa_\beta^2+C_x\kappa_\epsilon^2}\bigg),\label{subx-gbeta}
\end{align}
where $C_x\geq 0$ is a constant, which indicates $x_{ij}$ and $x_{ij}-G_{ij}\beta_{jk}$ are two subGau variables with subGau parameter $\sqrt{C_x(\bar{g}^2\kappa_\beta^2+\kappa_\epsilon^2)}$.\par

\subsection{Convergence Rate of $\hat{\mathbf\Sigma}_{\omega_k\omega_s}$}
We derive the covergence rate of the estimate of the estimation error covariance matrix $\hat{\mathbf\Sigma}_{\boldsymbol\omega}$. Recall that 
\begin{align}
\hat b_{jk}=\frac1{n_k}\sum_{i=1}^{n_k}F_{ij}X_{ik}.
\end{align}
We first investigate that 
\begin{align}
\Pr(|F_{ij}X_{ik}|\geq t)\leq\Pr(|F_{ij}|\geq \bar{f})+\Pr\bigg(|X_{ik}|\geq \frac{t}{\bar{f}}\bigg)\leq2\exp\bigg(-\frac{t^2}{C_x\bar{f}^2\bar{g}^2\kappa_\beta^2+C_x\bar{f}^2\kappa_\epsilon^2}\bigg).
\end{align}
Hence, $\hat b_{jk}$ is a subGau variable with subGau parameter $C_x\bar{f}\sqrt{\bar{g}^2\kappa_\beta^2+\kappa_\epsilon^2}$, and $\hat b_{jk}\hat b_{js}$ is a subExp variable with subExp parameter $C_x^2\bar{f}^2(\bar{g}^2\kappa_\beta^2+\kappa_\epsilon^2)$. By using Theorem 9.3 in \citet{fa},
\begin{align}
\Pr\bigg(\bigg|\frac{1}M\sum_{j=1}^M(\hat b_{jk}\hat b_{js}-\texttt{E}(\hat b_{jk}\hat b_{js}))\bigg|\geq\frac{2t}{\texttt{var}(\hat b_{jk})\texttt{var}(\hat b_{js})}\bigg)\leq2\exp(-c_MMt^2),
\end{align}
where $c_M$ is a constant and $t\leq \sqrt M/2$.

The next step is showing what $\texttt{var}(\hat b_{jk})$, $\texttt{var}(\hat b_{js})$, and $\texttt{E}(\hat b_{jk}\hat b_{js}))$ are. Specifically,
\begin{align}
\texttt{var}(\hat b_{jk})=\texttt{var}\bigg(\frac1{n_k}\sum_{i=1}^{n_k}F_{ij}X_{ik}\bigg)=\frac1{n_k^2}\texttt{var}(X_{ik})\texttt{var}(F_{ij})=\frac1{n_k^2},
\end{align}
since $\texttt{var}(X_{ik})=\texttt{var}(F_{ij})=1$. As for $\texttt{E}(\hat b_{jk}\hat b_{js}))$,
\begin{align}
\texttt{E}(\hat b_{jk}\hat b_{ls}))&=\texttt{E}\bigg\{\bigg(\frac1{n_k}\sum_{i=1}^{n_k}F_{ij}X_{ik}\bigg)\bigg(\frac1{n_k}\sum_{i=1}^{n_s}F_{ij}X_{is}\bigg)\bigg\}=\texttt{E}\bigg(\frac1{n_kn_s}\sum_{i=1}^{n_k}\sum_{t=1}^{n_s}F_{ij}X_{ik}F_{tj}X_{ts}\bigg)\notag\\
&=\texttt{E}\bigg[\texttt{E}\bigg(\frac1{n_kn_s}\sum_{i=1}^{n_k}\sum_{t=1}^{n_s}F_{ij}F_{tj}\bigg|X_{ik}X_{ts}\bigg)\bigg]=\texttt{E}\bigg(\frac1{n_kn_s}\sum_{i=1}^{n_k}\sum_{t=1}^{n_s}X_{ik}X_{ts}\bigg)\notag\\
&=\texttt{E}\bigg(\frac1{n_kn_s}\sum_{i\in\mathcal O_{ks}}X_{ik}X_{is}\bigg)=\frac{n_{ks}\texttt{cov}(X_{ik},X_{is})}{n_kn_s},
\end{align}
where $\mathcal{O}_{ks}$ is the set of overlapping samples in the $k$th and $s$th GWAS cohorts. This consistent with the results in our previous work \citet{lorincz-comi2022mrbee}.

\subsection{Convergence rate of $\hat{\Sigma}_{\beta_k\beta_s}$}
Then, we investigate the sub-Gaussianity of the GWAS estimates. First,
\begin{align}
\hat\beta_{jk}-\beta_{jk}&=\frac1{n_k}\sum_{i=1}^{n_k}(G_{ij}^2-1)\beta_{jk}+\frac1{n_k}\sum_{i=1}^{n_k}G_{ij}(X_{ik}-G_{ij}\beta_{jk})\notag\\
&=\omega_{jk}=\omega_{1jk}+\omega_{2jk}.
\end{align}
For $\omega_{1jk}$, we first note that $(G_{ij}^2-1)$ is a subGau variable bounded by $\bar{g}^2+1+2\bar{g}$. Hence,
\begin{align} 
\Pr(|(G_{ij}^2-1)\beta_{jk}|\geq t)\leq\Pr\bigg(|\beta_{jk}|\geq \frac{t}{\bar{g}^2+1+2\bar{g}}\bigg)\leq2\exp\bigg(-\frac{mt^2}{2\kappa_\beta^2(\bar{g}^2+1+2\bar{g})^2}\bigg),
\end{align}
and
\begin{align} 
\Pr(|\omega_{1jk}|\geq t)=\Pr\bigg(\bigg|\sum_{i=1}^{n_k}(G_{ij}^2-1)\beta_{jk}\bigg|\geq n_kt\bigg)\leq 2\exp\bigg(-\frac{n_kmt^2}{2\kappa_\beta^2(\bar{g}^2+1+2\bar{g})^2}\bigg),
\end{align}
which indicates that $\omega_{1jk}$ is a subGau variable with a subGau parameter $\sqrt{\frac{C_x}{n_km}\kappa_\beta(\bar{g}^2+1+2\bar{g})}$. The estimation error $\omega_{2jk}$ satisfies:
\begin{align} 
\Pr(|\omega_{2jk}|\geq t)=\Pr\bigg(\bigg|\sum_{i=1}^{n_k}G_{ij}(X_{ik}-G_{ij}\beta_{jk})\bigg|\geq n_kt\bigg)&\leq\Pr(|G_{ij}|\geq \bar{g})+\Pr\bigg(\bigg|\sum_{i=1}^{n_k}X_{ik}-G_{ij}\beta_{jk}\bigg|\geq \frac{n_kt}{\bar{g}}\bigg)\notag\\
&\leq2\exp\bigg(-\frac{n_kt^2}{C_x\bar{g}^4\kappa_\beta^2+C_x\kappa_\epsilon^2\bar{g}^2}\bigg),
\end{align}
which indicates that $\omega_{2jk}$ is a subGau variable with a subGau parameter $\sqrt{\frac{C_x}{n_k}(\bar{g}^4\kappa_\beta^2+\kappa_\epsilon^2\bar{g}^2)}$. Note that the subGau parameter of $\omega_{1jk}$ is smaller than $\omega_{2jk}$ when 
\begin{align}
m\geq \frac{\kappa_\beta^2(\bar{g}^2+1+2\bar{g})^2}{\bar{g}^4\kappa_\beta^2+\kappa_\epsilon^2\bar{g}^2}.
\label{mcondition}
\end{align}
Hence, we can consider $\omega_{jk}=\omega_{1jk}+\omega_{2jk}$ is a subGau variable with subGau parameter $\sqrt{\frac{2C_x}{n_k}(\bar{g}^4\kappa_\beta^2+\kappa_\epsilon^2\bar{g}^2)}$ if (\ref{mcondition}) holds.

We consider 
\begin{align}
\frac1m\sum_{j=1}^m\hat\beta_{jk}\hat\beta_{js}=\frac1m\sum_{j=1}^m(\beta_{jk}+\omega_{jk})(\beta_{js}+\omega_{js})=\frac1m\sum_{j=1}^m(\beta_{jk}\beta_{js}+\omega_{jk}\beta_{js}+\omega_{js}\beta_{jk}+\omega_{jk}\omega_{js}),
\end{align}
and hence
\begin{align}
\frac1m\sum_{j=1}^m(\hat\beta_{jk}\hat\beta_{js}-\Sigma_{w_kw_s}-\Sigma_{\beta_k\beta_s})=I_1+I_2+I_3+I_4,
\end{align}
where
\begin{align*}
I_1=\frac1m\sum_{j=1}^m(\beta_{jk}\beta_{js}-\Sigma_{\beta_k\beta_s}),~ I_2=\frac1m\sum_{j=1}^m(\omega_{jk}\omega_{js}-\Sigma_{\omega_k\omega_s}),~ I_3=\frac1m\sum_{j=1}^m\omega_{jk}\beta_{js},~ I_4=\frac1m\sum_{j=1}^m\omega_{js}\beta_{jk}.
\end{align*}
As for $I_1$, $\beta_{jk}\beta_{js}$ is a subExp. variable with a subExp. parameter $\frac1m\kappa_\beta^2$ that leads to 
\begin{align}
\Pr(|I_1|\geq t\sqrt{\Sigma_{\beta_k\beta_k}\Sigma_{\beta_s\beta_s}})\leq2\exp\bigg\{-mC_{\beta\beta}\min\bigg(\frac{m^2t^2\Sigma_{\beta_k\beta_k}\Sigma_{\beta_s\beta_s}}{\kappa_\beta^4},\frac{mt\sqrt{\Sigma_{\beta_k\beta_k}\Sigma_{\beta_s\beta_s}}}{\kappa_\beta^2}\bigg)\bigg\},
\label{I_1}
\end{align}
where $C_{\beta\beta}$ is a constant. 

As for $I_2$, $\omega_{jk}\omega_{js}$ is a subExp. variable with a subExp. parameter $\frac{C_{I_2}}{\sqrt{n_kn_s}}$ that leads to 
\begin{align}
\Pr(|I_2|\geq t\sqrt{\Sigma_{\beta_k\beta_k}\Sigma_{\beta_s\beta_s}})\leq2\exp\bigg\{-mC_{\omega\omega}\min\bigg(\frac{t^2n_kn_s\Sigma_{\beta_k\beta_k}\Sigma_{\beta_s\beta_s}}{C_{I_2}^2},\frac{t\sqrt{n_kn_s\Sigma_{\beta_k\beta_k}\Sigma_{\beta_s\beta_s}}}{C_{I_2}}\bigg)\bigg\},
\label{I_2}
\end{align}
where $C_{I_2}=2C_x(\bar{g}^4\kappa_\beta^2+\kappa_\epsilon^2\bar{g}^2)$ and $C_{\omega\omega}$ is a constant. 

As for $I_3$, $\omega_{jk}\beta_{js}$ is a subExp. variable with a subExp. parameter $\frac{C_{I3}}{\sqrt{n_km}}$ that leads to
\begin{align}
\Pr(|I_3|\geq t\sqrt{\Sigma_{\beta_k\beta_k}\Sigma_{\beta_s\beta_s}})\leq2\exp\bigg\{-mC_{\beta\omega}\min\bigg(\frac{t^2n_km\Sigma_{\beta_k\beta_k}\Sigma_{\beta_s\beta_s}}{C_{I_3}^2},\frac{t\sqrt{mn_k\Sigma_{\beta_k\beta_k}\Sigma_{\beta_s\beta_s}}}{C_{I_3}}\bigg)\bigg\},
\label{I_3}
\end{align}
where $C_{I_3}=2C_x(\bar{g}^4\kappa_\beta^4+\kappa_\epsilon^2\bar{g}^2\kappa_\beta^2)^{\frac12}$ and $C_{\beta\omega}$ is a constant. 

As for $I_4$, it is easy to obtain 
\begin{align}
\Pr(|I_4|\geq t\sqrt{\Sigma_{\beta_k\beta_k}\Sigma_{\beta_s\beta_s}})\leq2\exp\bigg\{-mC_{\beta\omega}\min\bigg(\frac{t^2n_sm\Sigma_{\beta_k\beta_k}\Sigma_{\beta_s\beta_s}}{C_{I_3}^2},\frac{t\sqrt{mn_s\Sigma_{\beta_k\beta_k}\Sigma_{\beta_s\beta_s}}}{C_{I_3}}\bigg)\bigg\},
\label{I_4}
\end{align}
since it is essentially a duplicate of $I_3$. Note that 
\begin{align}
\Pr\bigg\{\bigg|\frac1m\sum_{j=1}^m(\hat\beta_{jk}\hat\beta_{js}-\Sigma_{w_kw_s}-\Sigma_{\beta_k\beta_s})\bigg|\geq t\bigg\}\leq\sum_{l=1}^4\Pr(|I_l|\geq t),
\end{align}
and 
\begin{align}
\Sigma_{\beta_k\beta_k}\Sigma_{\beta_s\beta_s}=\frac{h^2_kh^2_s}{m^2},
\end{align}
$h^2_k$, $h^2_s$ are two constants. On the other, we apply a universal constant $\bar{C}$ to replace all constant in (\ref{I_1}-\ref{I_4}) and apply the minimum sample size $n_\textsf{min}$ to replace the sample size $n_k$, $n_s$, which results in
\begin{align}
&\Pr(|I_1|\geq t\sqrt{\Sigma_{\beta_k\beta_k}\Sigma_{\beta_s\beta_s}})\leq2\exp\bigg(-\bar{C}\min(t^2m,tm)\bigg),\\
&\Pr(|I_2|\geq t\sqrt{\Sigma_{\beta_k\beta_k}\Sigma_{\beta_s\beta_s}})\leq2\exp\bigg(-\bar{C}\min(t^2n_\textsf{min}^2m^{-1},tn_\textsf{min})\bigg),\\
\Pr(|I_3|\geq t\sqrt{\Sigma_{\beta_k\beta_k}\Sigma_{\beta_s\beta_s}})=&\Pr(|I_4|\geq t\sqrt{\Sigma_{\beta_k\beta_k}\Sigma_{\beta_s\beta_s}})\leq2\exp\bigg(-\bar{C}\min(t^2n_\textsf{min}m,t\sqrt{n_\textsf{min}m})\bigg).
\end{align}
Hence, by letting 
\begin{align}
n_\text{eff}=\min\{m,M,n_1,\dots,n_p\},
\end{align}
we prove that for any $t\leq 1$,
\begin{align}
\Pr\bigg\{\bigg|\frac1m\sum_{j=1}^m\bigg(\hat\beta_{jk}\hat\beta_{js}-\Sigma_{w_kw_s}-\Sigma_{\beta_k\beta_s}\bigg)\bigg|\geq t\sqrt{\Sigma_{\beta_k\beta_k}\Sigma_{\beta_s\beta_s}}\bigg\}\leq8\exp(-\bar{C}n_\text{eff}t^2).
\end{align}

After proving this central concentration inequality, we then to prove $\hat{\mathbf\Sigma}_{\boldsymbol\beta}^\texttt{Pearson}$ and $\hat{\mathbf\Sigma}_{\boldsymbol\beta}^\texttt{Spearman}$. For $\hat{\mathbf\Sigma}_{\boldsymbol\beta}^\texttt{Pearson}$,
\begin{align}
\hat\Sigma_{\beta_k\beta_s}^\texttt{Pearson}-\Sigma_{\beta_k\beta_s}=\frac1m\sum_{j=1}^m(\hat\beta_{jk}\hat\beta_{js}-\Sigma_{w_kw_s}-\Sigma_{\beta_k\beta_s})+\Sigma_{w_kw_s}-\hat\Sigma_{w_kw_s}.
\end{align}
Hence, 
\begin{align}
\Pr(|\hat\Sigma_{\beta_k\beta_s}^\texttt{Pearson}-\Sigma_{\beta_k\beta_s}|\geq t\sqrt{\Sigma_{\beta_k\beta_k}\Sigma_{\beta_s\beta_s}})&\leq\Pr\bigg\{\bigg|\frac1m\sum_{j=1}^m\bigg(\hat\beta_{jk}\hat\beta_{js}-\Sigma_{w_kw_s}-\Sigma_{\beta_k\beta_s}\bigg)\bigg|\geq t\sqrt{\Sigma_{\beta_k\beta_k}\Sigma_{\beta_s\beta_s}}\bigg)\notag\\
&+\Pr\bigg(|\Sigma_{w_kw_s}-\hat\Sigma_{w_kw_s}|\geq t\sqrt{\Sigma_{\beta_k\beta_k}\Sigma_{\beta_s\beta_s}}\bigg)\notag\\
&\leq10\exp(-\bar{C}n_\textsf{eff}t^2),
\end{align}
for any $t\leq 1$.

As for $\hat{\mathbf\Sigma}_{\boldsymbol\beta}^\texttt{Spearman}$, it has the same formula of central concentration inequality as $\hat{\mathbf\Sigma}_{\boldsymbol\beta}^\texttt{Pearson}$:
\begin{align}
\Pr(|\hat\Sigma_{\beta_k\beta_s}^\texttt{Spearman}-\Sigma_{\beta_k\beta_s}|\geq t\sqrt{\Sigma_{\beta_k\beta_k}\Sigma_{\beta_s\beta_s}})\leq C'\exp(-\bar{C}'n_\textsf{eff}t^2),
\end{align}
where $C',\bar{C}'$ is are two constants. Since $\hat\beta_{jk}=\beta_{jk}+\omega_{jk}$ is the sum of two subGau variable (and hence is also a subGau variable), we use the results of \citet{avella2018robust} to prove that 
\begin{align}
\Pr\bigg(|\hat D_{\beta_s}\hat D_{\beta_k}\hat R_{\beta_s\beta_k}-\texttt{se}(\hat\beta_{js})\texttt{cor}(\hat\beta_{jk}\hat\beta_{js})\texttt{se}(\hat\beta_{js})|\geq t\texttt{se}(\hat\beta_{jk})\texttt{se}(\hat\beta_{js})\bigg)\leq C_1\exp(-C_2mt^2),
\end{align}
where 
$$
\texttt{se}(\hat\beta_{jk})=\Sigma_{\beta_k\beta_k}+\Sigma_{\omega_k\omega_k},\quad \texttt{cor}(\hat\beta_{jk}\hat\beta_{js})=\frac{\Sigma_{\beta_k\beta_s}+\Sigma_{\omega_k\omega_s}}{\sqrt{\Sigma_{\beta_k\beta_k}+\Sigma_{\omega_k\omega_k}}\sqrt{\Sigma_{\beta_s\beta_s}+\Sigma_{\omega_s\omega_s}}},
$$
and hence 
$$
\texttt{se}(\hat\beta_{js})\texttt{cor}(\hat\beta_{jk}\hat\beta_{js})\texttt{se}(\hat\beta_{js})=\Sigma_{\beta_k\beta_s}+\Sigma_{\omega_k\omega_s}.
$$
and $C_1,C_2$ are certain constants. Therefore,
\begin{align}
\Pr(|\hat D_{\beta_s}\hat D_{\beta_k}\hat R_{\beta_s\beta_k}&-\Sigma_{\beta_k\beta_s}-\Sigma_{\omega_k\omega_s}+\Sigma_{\omega_k\omega_s}-\hat\Sigma_{\omega_k\omega_s}|\geq t\sqrt{\Sigma_{\beta_k\beta_k}\Sigma_{\beta_s\beta_s}})\notag\\
&\leq \Pr(|\hat D_{\beta_s}\hat D_{\beta_k}\hat R_{\beta_s\beta_k})-\Sigma_{\beta_k\beta_s}-\Sigma_{\omega_k\omega_s}|\geq t\sqrt{\Sigma_{\beta_k\beta_k}\Sigma_{\beta_s\beta_s}})\notag\\
&+\Pr(|\Sigma_{\omega_k\omega_s}-\hat\Sigma_{\omega_k\omega_s}|\geq t\sqrt{\Sigma_{\beta_k\beta_k}\Sigma_{\beta_s\beta_s}}).
\end{align}
Here, 
\begin{align}
\Pr(|\hat D_{\beta_s}\hat D_{\beta_k}\hat R_{\beta_s\beta_k}&-\Sigma_{\beta_k\beta_s}-\Sigma_{\omega_k\omega_s}|\geq t\sqrt{\Sigma_{\beta_k\beta_k}\Sigma_{\beta_s\beta_s}})\notag\\
&=\Pr(|\hat D_{\beta_s}\hat D_{\beta_k}\hat R_{\beta_s\beta_k})-\Sigma_{\beta_k\beta_s}-\Sigma_{\omega_k\omega_s}|\geq t'\texttt{se}(\hat\beta_{jk})\texttt{se}(\hat\beta_{js})),
\end{align}
where 
$$
t'=t\sqrt{\frac{\Sigma_{\beta_k\beta_k}\Sigma_{\beta_s\beta_s}}{(\Sigma_{\beta_k\beta_k}+\Sigma_{\omega_k\omega_k})(\Sigma_{\beta_s\beta_s}+\Sigma_{\omega_s\omega_s})}}.
$$
Note that $t'<1$ because $\Sigma_{\beta_k\beta_k}, \Sigma_{\omega_k\omega_k},\Sigma_{\beta_s\beta_s},\Sigma_{\omega_s\omega_s}$ are all positive numbers. Hence. by choosing $C',\bar{C}'$ constants properly, we prove 
\begin{align}
\Pr(|\hat\Sigma_{\beta_k\beta_s}^\texttt{Spearman}-\Sigma_{\beta_k\beta_s}|\geq t\sqrt{\Sigma_{\beta_k\beta_k}\Sigma_{\beta_s\beta_s}})\leq C'\exp(-\bar{C}'n_\textsf{eff}t^2).
\end{align}

\subsection{Proof of Theorem 1}
It is easy to see that 
\begin{align}
\Pr\bigg\{\max_{1\leq k\leq s\leq p}\bigg(|\hat\Sigma_{\beta_k\beta_s}^\texttt{Pearson}-\Sigma_{\beta_k\beta_s}|\geq  t\sqrt{\Sigma_{\beta_k\beta_k}\Sigma_{\beta_s\beta_s}}\bigg)\bigg\}&\leq\sum_{j=1}^p\sum_{s=1}^p\Pr\bigg(|\hat\Sigma_{\beta_k\beta_s}^\texttt{Pearson}-\Sigma_{\beta_k\beta_s}|\geq  t\sqrt{\Sigma_{\beta_k\beta_k}\Sigma_{\beta_s\beta_s}}\bigg)\notag\\
\leq 10p^2\exp(-\bar{C}n_\texttt{eff}t^2)=\exp(\log(10p^2)-\bar{C}n_\texttt{eff}t^2),
\end{align}
\begin{align}
\Pr\bigg\{\max_{1\leq k\leq s\leq p}\bigg(|\hat\Sigma_{\beta_k\beta_s}^\texttt{Spearman}-\Sigma_{\beta_k\beta_s}|\geq  t\sqrt{\Sigma_{\beta_k\beta_k}\Sigma_{\beta_s\beta_s}}\bigg)\bigg\}&\leq\sum_{j=1}^p\sum_{s=1}^p\Pr\bigg(|\hat\Sigma_{\beta_k\beta_s}^\texttt{Spearman}-\Sigma_{\beta_k\beta_s}|\geq  t\sqrt{\Sigma_{\beta_k\beta_k}\Sigma_{\beta_s\beta_s}}\bigg)\notag\\
\leq C'p^2\exp(-\bar{C}'n_\texttt{eff}t^2)=\exp(\log(C'p^2)-\bar{C}'n_\texttt{eff}t^2).
\end{align}
herefore, by choosing $c_{\Sigma_1}=\max(10,C')$ and $c_{\Sigma_2}=\max(\bar{C},\bar{C}')$, we prove
\[
\Pr\bigg(\max_{1\leq j\leq k\leq p}|\hat\Sigma_{\beta_j\beta_k}-\Sigma_{\beta_j\beta_k}|\leq t\sqrt{\Sigma_{\beta_j\beta_j}\Sigma_{\beta_k\beta_k}}\bigg)\geq1-\exp\bigg(\log(c_{\Sigma_1}p^2)-n_\texttt{eff}c_{\Sigma_2} c_{\Sigma_2} t^2\bigg)\]
where $\hat\Sigma_{\beta_j\beta_k}$ can be $\hat\Sigma_{\beta_k\beta_s}^\texttt{Pearson}$ or $\hat\Sigma_{\beta_k\beta_s}^\texttt{Spearman}$.

\subsection{Proof of Theorem 2}
We apply the primal-dual witness(PDW) approach \citep{ravikumar2011high} to proof Theorem 3.2, which consists of three steps:
\begin{enumerate}
\item[Step 1] Solve an oracle minimization:
\begin{align}
\hat{\mathbf\Theta}_{\boldsymbol\beta}^\texttt{oracle}=\arg\min_{\mathbf\Theta\in\mathcal{O}_{\mathbf\Theta}}\bigg\{
\texttt{entropy}(\hat{\mathbf\Sigma}_{\boldsymbol\theta},\mathbf\Theta)+\sum_{k=1}^p\sum_{s\neq k}P_{\lambda}(|\Theta_{\beta_k\beta_s}|)
\bigg\},
\end{align}
where
\[
\mathcal{O}_{\mathbf\Theta}=\bigg\{\mathbf\Theta:\ \Theta_{ks}=0\text{ if }(k,j)\in\mathcal{S},\sigma_{\min }(\mathbf\Theta)\geq\delta\bigg\}.
\]
\item[Step 2] Prove
\[
\Pr\bigg(|\hat\Theta_{\beta_k\beta_s}^\texttt{oracle}-\Theta_{\beta_k\beta_s}|\geq t\sqrt{\Theta_{\beta_j\beta_j}\Theta_{\beta_k\beta_k}}\bigg)\geq1-\exp(\log(c_{\Theta_1}p^2)-n_\texttt{eff}c_{\Theta_2} c_{\Sigma_2} t^2),
\]
for all $(j,k)\in\mathcal{S}$.
\item[Step 3] Verify if $\lambda=c_\lambda(\max_{1\leq j \leq p}\Theta_{\beta_j\beta_j})t$ and $\min_{1\leq k\leq j\leq p}|\Theta_{jk}|/\lambda\to\infty$,
\[
\Pr\bigg(\hat P_{ks}<P'_\lambda(0)\bigg)\geq1-\exp(\log(c_{\Theta_1}p^2)-n_\texttt{eff}c_{\Theta_2} c_{\Sigma_2} t^2),
\]
for all $(j,k)\not\in\mathcal{S}$, where 
\begin{align}
\hat P_{ks}=\partial P_\lambda(|\hat\Theta_{\beta_k\beta_s}^\texttt{oracle}|)/\partial \Theta_{\beta_k\beta_s}.
\end{align}
\end{enumerate}
By combing Steps 2 and 3, we can obtain 
$$
\Pr\bigg(\forall(j,k)\in\mathcal S,\ \texttt{sign}(\hat\Theta_{\beta_j\beta_k})=\texttt{sign}(\Theta_{\beta_j\beta_k})\bigg)\geq1-\exp(\log(c_{\Theta_1}p^2)-n_\texttt{eff}c_{\Theta_2} c_{\Sigma_2} t^2).
$$
Note that 
\[
\mathcal S=\{(k,s),\Theta_{\beta_k\beta_s}\neq0\}.
\]

We first work with the KKT condition of Step1:
\begin{align}
\hat{\mathbf\Sigma}_{\boldsymbol\beta}-(\hat{\mathbf\Theta}_{\boldsymbol\beta}^\texttt{oracle})^{-1}+\hat{\mathbf P}=\hat{\mathbf\Sigma}_{\boldsymbol\beta}-\mathbf\Sigma_{\boldsymbol\beta}+\{\mathbf\Theta^{-1}_{\boldsymbol\beta}-(\hat{\mathbf\Theta}_{\boldsymbol\beta}^\texttt{oracle})^{-1}\}+\hat{\mathbf P}&=\mathbf0\notag\\
\Rightarrow\hat{\mathbf W}_{\boldsymbol\beta}+\{\mathbf\Theta_{\boldsymbol\beta}^{-1}-(\hat{\mathbf\Theta}_{\boldsymbol\beta}^\texttt{oracle})^{-1}\}+\hat{\mathbf P}&=\mathbf0,
\end{align}
where $\hat{\mathbf W}=\hat{\mathbf\Sigma}_{\boldsymbol\beta}-\mathbf\Sigma_{\boldsymbol\beta}$ and the $(k,s)$th element of $\hat{\mathbf P}$ is $\hat P_{ks}$. Let 
$$
\hat{\boldsymbol w}=\texttt{vec}(\hat{\mathbf W}),\quad \hat{\boldsymbol\theta}^\texttt{oracle}_{\boldsymbol\beta}=\texttt{vec}(\hat{\mathbf\Theta}_{\boldsymbol\beta}^\texttt{oracle}),\quad \hat{\boldsymbol p}=\texttt{vec}(\hat{\mathbf P}).
$$
Using the Taylor expression at $\boldsymbol\theta_{\boldsymbol\beta}=\texttt{vec}(\mathbf\Theta_{\boldsymbol\beta})$, we turn the KKT condition into
\begin{align}
\begin{pmatrix}
\hat{\boldsymbol w}_{\mathcal S}\\
\hat{\boldsymbol w}_{\bar{\mathcal S}}
\end{pmatrix}
-
\begin{pmatrix}
\mathbf\Upsilon_{\mathcal S\mathcal S}&\mathbf\Upsilon_{\mathcal S\bar{\mathcal S}}\\
\mathbf\Upsilon_{\bar{\mathcal S}\mathcal S}&\mathbf\Upsilon_{\bar{\mathcal S}\bar{\mathcal S}}
\end{pmatrix}
\begin{pmatrix}
\hat{\boldsymbol\theta}_{\mathcal S}-\boldsymbol\theta_{\mathcal S}\\
\mathbf 0
\end{pmatrix}
+
\begin{pmatrix}
\hat{\boldsymbol p}_{\mathcal S}\\
\hat{\boldsymbol p}_{\bar{\mathcal S}}
\end{pmatrix}-
\begin{pmatrix}
\boldsymbol r_{\mathcal S}\\
\boldsymbol r_{\bar{\mathcal S}}
\end{pmatrix}
=\mathbf 0.
\end{align}
where $\boldsymbol r=(\boldsymbol r_{\mathcal S}^\top,\boldsymbol r_{\bar{\mathcal S}}^\top)^\top$ is the residual vector of the Taylor expression, whose expression is
\begin{align}
\boldsymbol r=\texttt{vec}\bigg((\hat{\mathbf\Theta}^{\texttt{oracle}})^{-1}-\mathbf\Theta_{\boldsymbol\beta}^{-1}+\mathbf\Theta_{\boldsymbol\beta}^{-1}(\hat{\mathbf\Theta}^{\texttt{oracle}}-\mathbf\Theta_{\boldsymbol\beta})\mathbf\Theta_{\boldsymbol\beta}^{-1}\bigg),
\end{align}
according to the conclusion in \citet[Equation (50)]{ravikumar2011high}, and we remove the symbol $\boldsymbol\beta$ in $\hat{\boldsymbol w}_{\mathcal S}$,
$\hat{\boldsymbol w}_{\bar{\mathcal S}}$, $\hat{\boldsymbol\theta}^\texttt{oracle}_{\mathcal S}$, $\hat{\boldsymbol\theta}^\texttt{oracle}_{\bar{\mathcal S}}$ to simplify the notations. Thus, we obtain two equations for $\hat{\boldsymbol\theta}^\texttt{oracle}_{\mathcal S}$ and $\hat{\boldsymbol\theta}^\texttt{oracle}_{\bar{\mathcal S}}$ (which is a zero vector):
\begin{align}
\hat{\boldsymbol w}_{\mathcal S}-\mathbf\Upsilon_{\mathcal S\mathcal S}(\hat{\boldsymbol\theta}_{\mathcal S}-\boldsymbol\theta_{\mathcal S})+\hat{\boldsymbol p}_{\mathcal S}-\boldsymbol r_{\mathcal S}=\mathbf 0\\
\hat{\boldsymbol w}_{\bar{\mathcal S}}-\mathbf\Upsilon_{\bar{\mathcal S}\mathcal S}(\hat{\boldsymbol\theta}_{\mathcal S}-\boldsymbol\theta_{\mathcal S})+\hat{\boldsymbol p}_{\bar{\mathcal S}}-\boldsymbol r_{\bar{\mathcal S}}=\mathbf0.
\end{align}

We first work with the first equation. By combining condition (C6) (i.e., $P'_\lambda(|x|)=0$ for $x\in[a,+\infty)$ where $a$ is a constant) and $\min_{1\leq k\leq j\leq p}|\Theta_{jk}|/\lambda\to\infty$, we obtain
\begin{align}
\hat{\boldsymbol p}_{\mathcal S}=\mathbf 0.
\end{align}
Thus,
\begin{align}
||\mathbf\Upsilon_{\mathcal S\mathcal S}(\hat{\boldsymbol\theta}_{\mathcal S}-\boldsymbol\theta_{\mathcal S})||_\infty\leq||\hat{\boldsymbol w}_{\mathcal S}||_\infty+||\boldsymbol r||_\infty\Rightarrow||\mathbf\Upsilon_{\mathcal S\mathcal S}||_\infty||\hat{\boldsymbol\theta}_{\mathcal S}-\boldsymbol\theta_{\mathcal S}||\leq ||\hat{\boldsymbol w}_{\mathcal S}||_\infty+||\boldsymbol r||_\infty.
\end{align}
By using \citep{ravikumar2011high}[Lemma 4], there exist a constant $D_1$ such that
\[
||\boldsymbol r||_\infty\leq D_1||\hat{\boldsymbol w}||_\infty.
\]
Therefore,
\begin{align}
||\hat{\boldsymbol\theta}_{\mathcal S}-\boldsymbol\theta_{\mathcal S})||_\infty\leq\frac{D_1}{||\mathbf\Upsilon_{\mathcal S\mathcal S}||_\infty}||\hat{\boldsymbol w}||_\infty,
\end{align}
and hence
\begin{align}
\Pr\bigg(||\hat{\boldsymbol\theta}_{\mathcal S}-\boldsymbol\theta_{\mathcal S})||_\infty\geq t\sqrt{\Theta_{\beta_k\beta_k}\Theta_{\beta_s\beta_s}}\bigg)\leq \Pr\bigg(
||\hat{\boldsymbol w}_{\mathcal S}||_\infty\geq \frac{t||\mathbf\Upsilon_{\mathcal S\mathcal S}||_\infty\sqrt{\Theta_{\beta_k\beta_k}\Theta_{\beta_s\beta_s}}}{D_1}
\bigg).
\end{align}
Note that 
\begin{align}
||\mathbf\Upsilon_{\mathcal S\mathcal S}||_\infty=\max_{i,j,k,s}\bigg\{|\Sigma_{\beta_i\beta_j}\Sigma_{\beta_k\beta_s}|\bigg\},
\end{align}
and hence
\begin{align}
||\mathbf\Upsilon_{\mathcal S\mathcal S}||_\infty\sqrt{\Theta_{\beta_k\beta_k}\Theta_{\beta_s\beta_s}}\leq \max_{k,s}\sqrt{\Sigma_{\beta_k\beta_k}\Sigma_{\beta_s\beta_s}}.
\end{align}
Now we move to Step 3. It is easy to see that
\begin{align}
||\hat{\boldsymbol p}_{\bar{\mathcal S}}||_\infty&\leq||\mathbf{\Upsilon}_{\bar{\mathcal S}\mathcal S}(\hat{\boldsymbol\theta}_{\mathcal S}-\boldsymbol\theta_{\mathcal S})||_\infty+||\hat{\boldsymbol w}_{\bar{\mathcal S}}||_\infty+||\boldsymbol r_{\bar{\mathcal S}}||_\infty\notag\\
&\leq||\mathbf{\Upsilon}_{\bar{\mathcal S}\mathcal S}\mathbf{\Upsilon}_{\mathcal S\mathcal S}^{-1}(\hat{\boldsymbol w}_\mathcal{S}-\boldsymbol r_\mathcal{S})||_\infty+||\hat{\boldsymbol w}_{\bar{\mathcal S}}||_\infty+||\boldsymbol r_{\bar{\mathcal S}}||_\infty\notag\\
&\leq(||\mathbf{\Upsilon}_{\bar{\mathcal S}\mathcal S}\mathbf{\Upsilon}_{\mathcal S\mathcal S}^{-1}||_\infty+1)(D_1+1)||\hat{\boldsymbol w}||_\infty\leq (c_\upsilon+1)(D_1+1)||\hat{\boldsymbol w}||_\infty,
\end{align}
where $c_\upsilon\geq ||\mathbf{\Upsilon}_{\bar{\mathcal S}\mathcal S}\mathbf{\Upsilon}_{\mathcal S\mathcal S}^{-1}||_\infty$. By condition (C6), 
$$
P'_\lambda(0)\leq a_1\lambda
$$
with a constant $a_1.$ Subsequently,
\begin{align}
\frac{||\hat{\boldsymbol p}_{\bar{\mathcal S}}||_\infty}{P'_\lambda(0)}\leq \frac{ (c_\upsilon+1)(D_1+1)||\hat{\boldsymbol w}||_\infty}{a_1\lambda}.
\end{align}
By choosing $c_{\lambda_1}$ large enough,
\begin{align}
\Pr\bigg(||\hat{\boldsymbol\theta}_{\mathcal S}-\boldsymbol\theta_{\mathcal S}||_\infty\geq t\sqrt{\Theta_{\beta_k\beta_k}\Theta_{\beta_s\beta_s}}\bigg)&\leq \Pr\bigg(
||\hat{\boldsymbol w}_{\mathcal S}||_\infty\geq \frac{t\sqrt{\Sigma_{\beta_k\beta_k}\Sigma_{\beta_s\beta_s}}}{D_1}
\bigg)\notag\\
&\leq c_{\Sigma_1}\exp(-D_1^{-2}c_{\Sigma_2}n_\texttt{eff}t^2)\leq c_{\Theta_1}\exp(-c_{\Theta_2}n_\texttt{eff}t^2),
\end{align}
\begin{align}
\Pr\bigg(\frac{||\hat{\boldsymbol p}_{\bar{\mathcal S}}||_\infty}{P'_\lambda(0)}\geq1\bigg)&\leq\Pr\bigg(\frac{(c_\upsilon+1)(D_1+1)||\hat{\boldsymbol w}||_\infty}{a_1\lambda}\geq t\bigg)=\Pr\bigg(||\hat{\boldsymbol w}||_\infty\geq\frac{a_1\lambda t}{(c_\upsilon+1)(D_1+1)}\bigg)\notag\\
&\leq c_{\Sigma_1}\exp\bigg(-\frac{c_{\Sigma_2}a_1^2c_{\lambda_1}^2n_\texttt{eff}t^2}{(c_\upsilon+1)(D_1+1)}\bigg),
\end{align}
Note that 
\begin{align}
\hat\Theta_{\beta_k\beta_s}-\Theta_{\beta_k\beta_s}=\mathbf\Upsilon_{\mathcal S\mathcal S}^{-1}(\hat{\boldsymbol w}_{\mathcal S}-\boldsymbol r_{\mathcal X}).
\end{align}
Therefore,
\begin{align}
\Pr\bigg(\forall(j,k)\in\mathcal S,\ \texttt{sign}(\hat\Theta_{\beta_j\beta_k})&\neq\texttt{sign}(\Theta_{\beta_j\beta_k})\bigg)\leq \Pr\bigg(\sigma_{\min}(\mathbf\Upsilon_{\mathcal S\mathcal S}^{-1})(D_1+1)||\hat{\boldsymbol w}_{\mathcal S}||_\infty\geq|\Theta_{\beta_k\beta_s}|\bigg)\notag\\
&\leq\Pr\bigg(\sigma_{\min}(\mathbf\Upsilon_{\mathcal S\mathcal S}^{-1})(D_1+1)||\hat{\boldsymbol w}_{\mathcal S}||_\infty\geq c_{\lambda_2}(\max_{1\leq k\leq p}\Theta_{\beta_k\beta_k})t)\bigg)\notag\\
&\leq c_{\Sigma_1}\exp\bigg(-\frac{c_{\lambda_2}^2c_{\Sigma_1}n_\texttt{eff}t^2}{(D_1+1)^2}\bigg).
\end{align}
Therefore, by choosing $c_{\Theta_1}$ and $c_{\Theta_2}$ large enough,
we prove that 
\[
\Pr\bigg(\forall(j,k),\ \texttt{sign}(\hat\Theta_{\beta_j\beta_k})=\texttt{sign}(\Theta_{\beta_j\beta_k})\bigg)\geq1-\exp\bigg(\log(c_{\Theta_1}p^2)-n_\texttt{eff}c_{\Theta_2} t^2\bigg),
\]
which implies
\[
\Pr\bigg(\max_{1\leq j\leq k\leq p}|\hat\Theta_{\beta_j\beta_k}-\Theta_{\beta_j\beta_k}|\leq t\sqrt{\Theta_{\beta_j\beta_j}\Theta_{\beta_k\beta_k}}\bigg)\geq1-\exp\bigg(\log(c_{\Theta_1}p^2)-n_\texttt{eff}c_{\Theta_2} t^2\bigg).\]

\section*{Data Availability}
The GWAS data in the Million Veteran Program (MVP) are available through dbGAP under accession number phs001672.v7.p1 (Veterans Administration MVP Summary Results from Omics Studies). Other GWAS data are available through the Data Availability section in the corresponding papers.

\section*{Acknowledgments}
We would like to thank the editor and the anonymous referees for their very careful reviews and constructive suggestions. This work was supported by grants HG011052 and HG011052-03S1 (to X.Z.) from the National Human Genome Research Institute (NHGRI), USA.

\bibliographystyle{abbrvnat}
\bibliography{refs.bib}

\begin{thebibliography}{72}
\providecommand{\natexlab}[1]{#1}
\providecommand{\url}[1]{\texttt{#1}}
\expandafter\ifx\csname urlstyle\endcsname\relax
  \providecommand{\doi}[1]{doi: #1}\else
  \providecommand{\doi}{doi: \begingroup \urlstyle{rm}\Url}\fi

\bibitem[Abdellaoui et~al.(2023)Abdellaoui, Yengo, Verweij, and
  Visscher]{abdellaoui202315}
A.~Abdellaoui, L.~Yengo, K.~J. Verweij, and P.~M. Visscher.
\newblock 15 years of gwas discovery: Realizing the promise.
\newblock \emph{AJHG}, 2023.

\bibitem[Ahmad et~al.(2015)Ahmad, Morris, Mujammami, Forgetta, Leong, Li,
  Turgeon, Greenwood, Thanassoulis, Meigs, et~al.]{ahmad2015mendelian}
O.~S. Ahmad, J.~A. Morris, M.~Mujammami, V.~Forgetta, A.~Leong, R.~Li,
  M.~Turgeon, C.~M. Greenwood, G.~Thanassoulis, J.~B. Meigs, et~al.
\newblock A mendelian randomization study of the effect of type-2 diabetes on
  coronary heart disease.
\newblock \emph{Nat. Commun.}, 6\penalty0 (1):\penalty0 7060, 2015.

\bibitem[Aragam et~al.(2022)]{aragam2022discovery}
K.~G. Aragam et~al.
\newblock Discovery and systematic characterization of risk variants and genes
  for coronary artery disease in over a million participants.
\newblock \emph{Nat. Genet.}, pages 1--13, 2022.

\bibitem[Avella-Medina et~al.(2018)Avella-Medina, Battey, Fan, and
  Li]{avella2018robust}
M.~Avella-Medina, H.~S. Battey, J.~Fan, and Q.~Li.
\newblock Robust estimation of high-dimensional covariance and precision
  matrices.
\newblock \emph{Biometrika}, 105\penalty0 (2):\penalty0 271--284, 2018.

\bibitem[Bowden et~al.(2015)Bowden, Davey~Smith, and
  Burgess]{bowden2015mendelian}
J.~Bowden, G.~Davey~Smith, and S.~Burgess.
\newblock Mendelian randomization with invalid instruments: effect estimation
  and bias detection through egger regression.
\newblock \emph{Int. J. Epidemiol.}, 44\penalty0 (2):\penalty0 512--525, 2015.

\bibitem[Bowden et~al.(2016)Bowden, Davey~Smith, Haycock, and
  Burgess]{bowden2016consistent}
J.~Bowden, G.~Davey~Smith, P.~C. Haycock, and S.~Burgess.
\newblock Consistent estimation in mendelian randomization with some invalid
  instruments using a weighted median estimator.
\newblock \emph{Genet. Epidemiol.}, 40\penalty0 (4):\penalty0 304--314, 2016.

\bibitem[Boyd et~al.(2011)Boyd, Parikh, Chu, Peleato, and
  Eckstein]{boyd2011distributed}
S.~Boyd, N.~Parikh, E.~Chu, B.~Peleato, and J.~Eckstein.
\newblock Distributed optimization and statistical learning via the alternating
  direction method of multipliers.
\newblock \emph{Found. Trends Mach. Learn.}, 3\penalty0 (1):\penalty0 1--122,
  2011.

\bibitem[B{\"u}hlmann and Van De~Geer(2011)]{buhlmann2011statistics}
P.~B{\"u}hlmann and S.~Van De~Geer.
\newblock \emph{Statistics for high-dimensional data: methods, theory and
  applications}.
\newblock Springer Science \& Business Media, 2011.

\bibitem[Bulik-Sullivan et~al.(2015{\natexlab{a}})]{bulik2015atlas}
B.~Bulik-Sullivan et~al.
\newblock An atlas of genetic correlations across human diseases and traits.
\newblock \emph{Nat. Genet.}, 47\penalty0 (11):\penalty0 1236--1241,
  2015{\natexlab{a}}.

\bibitem[Bulik-Sullivan et~al.(2015{\natexlab{b}})]{bulik2015ld}
B.~K. Bulik-Sullivan et~al.
\newblock Ld score regression distinguishes confounding from polygenicity in
  genome-wide association studies.
\newblock \emph{Nat. Genet.}, 47\penalty0 (3):\penalty0 291--295,
  2015{\natexlab{b}}.

\bibitem[Burgess et~al.(2013)Burgess, Butterworth, and
  Thompson]{burgess2013mendelian}
S.~Burgess, A.~Butterworth, and S.~G. Thompson.
\newblock Mendelian randomization analysis with multiple genetic variants using
  summarized data.
\newblock \emph{Genet. Epidemiol.}, 37\penalty0 (7):\penalty0 658--665, 2013.

\bibitem[Cai et~al.(2011)Cai, Liu, and Luo]{cai2011constrained}
T.~Cai, W.~Liu, and X.~Luo.
\newblock A constrained $\ell_1$ minimization approach to sparse precision
  matrix estimation.
\newblock \emph{J. Am. Stat. Assoc.}, 106\penalty0 (494):\penalty0 594--607,
  2011.

\bibitem[Chen et~al.(2020)]{chen2020trans}
M.-H. Chen et~al.
\newblock Trans-ethnic and ancestry-specific blood-cell genetics in 746,667
  individuals from 5 global populations.
\newblock \emph{Cell}, 182\penalty0 (5):\penalty0 1198--1213, 2020.

\bibitem[Cole et~al.(1990)Cole, Drummond, Osborne, and
  Matsumura]{cole1990hypertension}
D.~J. Cole, J.~C. Drummond, T.~N. Osborne, and J.~Matsumura.
\newblock Hypertension and hemodilution during cerebral ischemia reduce brain
  injury and edema.
\newblock \emph{Am. J. Physiol. Heart. Circ. Physiol.}, 259\penalty0
  (1):\penalty0 H211--H217, 1990.

\bibitem[Consortium(2010)]{international2010integrating}
I.~H.~. Consortium.
\newblock Integrating common and rare genetic variation in diverse human
  populations.
\newblock \emph{Nature}, 467\penalty0 (7311):\penalty0 52, 2010.

\bibitem[Fan and Li(2001)]{fan2001variable}
J.~Fan and R.~Li.
\newblock Variable selection via nonconcave penalized likelihood and its oracle
  properties.
\newblock \emph{J. Am. Stat. Assoc.}, 96\penalty0 (456):\penalty0 1348--1360,
  2001.

\bibitem[Fan et~al.(2014)Fan, Xue, and Zou]{fan2014strong}
J.~Fan, L.~Xue, and H.~Zou.
\newblock Strong oracle optimality of folded concave penalized estimation.
\newblock \emph{Ann. Stat.}, 42\penalty0 (3):\penalty0 819, 2014.

\bibitem[Feizi et~al.(2013)Feizi, Marbach, M{\'e}dard, and
  Kellis]{feizi2013network}
S.~Feizi, D.~Marbach, M.~M{\'e}dard, and M.~Kellis.
\newblock Network deconvolution as a general method to distinguish direct
  dependencies in networks.
\newblock \emph{Nat. Biotechnol.}, 31\penalty0 (8):\penalty0 726--733, 2013.

\bibitem[Friedman et~al.(2008)Friedman, Hastie, and
  Tibshirani]{friedman2008sparse}
J.~Friedman, T.~Hastie, and R.~Tibshirani.
\newblock Sparse inverse covariance estimation with the graphical lasso.
\newblock \emph{Biostatistics}, 9\penalty0 (3):\penalty0 432--441, 2008.

\bibitem[Graham et~al.(2021)]{graham2021power}
S.~E. Graham et~al.
\newblock The power of genetic diversity in genome-wide association studies of
  lipids.
\newblock \emph{Nature}, 600\penalty0 (7890):\penalty0 675--679, 2021.

\bibitem[Group(2017)]{hps32017effects}
H.-R.~C. Group.
\newblock Effects of anacetrapib in patients with atherosclerotic vascular
  disease.
\newblock \emph{NEJM}, 377\penalty0 (13):\penalty0 1217--1227, 2017.

\bibitem[Ishigaki et~al.(2020)]{ishigaki2020large}
K.~Ishigaki et~al.
\newblock Large-scale genome-wide association study in a japanese population
  identifies novel susceptibility loci across different diseases.
\newblock \emph{Nat. Genet.}, 52\penalty0 (7):\penalty0 669--679, 2020.

\bibitem[Kanai et~al.(2018)]{kanai2018genetic}
M.~Kanai et~al.
\newblock Genetic analysis of quantitative traits in the japanese population
  links cell types to complex human diseases.
\newblock \emph{Nat. Genet.}, 50\penalty0 (3):\penalty0 390--400, 2018.

\bibitem[Kim et~al.(2022)]{kim2022contribution}
Y.~J. Kim et~al.
\newblock The contribution of common and rare genetic variants to variation in
  metabolic traits in 288,137 east asians.
\newblock \emph{Nat. Commun.}, 13\penalty0 (1):\penalty0 6642, 2022.

\bibitem[Langfelder and Horvath(2008)]{langfelder2008wgcna}
P.~Langfelder and S.~Horvath.
\newblock Wgcna: an r package for weighted correlation network analysis.
\newblock \emph{BMC Bioinf.}, 9\penalty0 (1):\penalty0 1--13, 2008.

\bibitem[Lauritzen(1996)]{lauritzen1996graphical}
S.~L. Lauritzen.
\newblock \emph{Graphical models}, volume~17.
\newblock Clarendon Press, 1996.

\bibitem[Le~Sueur et~al.(2020)Le~Sueur, Bruce, Geifman, and
  Consortium]{le2020challenges}
H.~Le~Sueur, I.~N. Bruce, N.~Geifman, and M.~Consortium.
\newblock The challenges in data integration--heterogeneity and complexity in
  clinical trials and patient registries of systemic lupus erythematosus.
\newblock \emph{BMC Med. Res. Methodol.}, 20:\penalty0 1--5, 2020.

\bibitem[Lin et~al.(2023)Lin, Xue, and Pan]{lin2023combining}
Z.~Lin, H.~Xue, and W.~Pan.
\newblock Combining mendelian randomization and network deconvolution for
  inference of causal networks with gwas summary data.
\newblock \emph{PLoS Genet.}, 19\penalty0 (5):\penalty0 e1010762, 2023.

\bibitem[Loh et~al.(2018)Loh, Kichaev, Gazal, Schoech, and Price]{loh2018mixed}
P.-R. Loh, G.~Kichaev, S.~Gazal, A.~P. Schoech, and A.~L. Price.
\newblock Mixed-model association for biobank-scale datasets.
\newblock \emph{Nat. Genet.}, 50\penalty0 (7):\penalty0 906--908, 2018.

\bibitem[Lorincz-Comi et~al.(2023)Lorincz-Comi, Yang, Li, and
  Zhu]{lorincz-comi2022mrbee}
N.~Lorincz-Comi, Y.~Yang, G.~Li, and X.~Zhu.
\newblock Mrbee: A novel bias-corrected multivariable mendelian randomization
  method.
\newblock \emph{biorxiv, 523480}, 2023.

\bibitem[Lu et~al.(2023)Lu, Zheng, and Quinn]{lu2023introducing}
Y.~Lu, Q.~Zheng, and D.~Quinn.
\newblock Introducing causal inference using bayesian networks and do-calculus.
\newblock \emph{J. Stat. Data Sci. Educ.}, 31\penalty0 (1):\penalty0 3--17,
  2023.

\bibitem[MacArthur et~al.(2017)]{macarthur2017new}
J.~MacArthur et~al.
\newblock The new nhgri-ebi catalog of published genome-wide association
  studies (gwas catalog).
\newblock \emph{Nucleic Acids Res.}, 45\penalty0 (D1):\penalty0 D896--D901,
  2017.

\bibitem[Mailman et~al.(2007)]{mailman2007ncbi}
M.~D. Mailman et~al.
\newblock The ncbi dbgap database of genotypes and phenotypes.
\newblock \emph{Nat. Genet.}, 39\penalty0 (10):\penalty0 1181--1186, 2007.

\bibitem[Meinshausen and B{\"u}hlmann(2006)]{meinshausen2006high}
N.~Meinshausen and P.~B{\"u}hlmann.
\newblock High-dimensional graphs and variable selection with the lasso.
\newblock \emph{Ann. Stat.}, pages 1436--1462, 2006.

\bibitem[Meinshausen and B{\"u}hlmann(2010)]{meinshausen2010stability}
N.~Meinshausen and P.~B{\"u}hlmann.
\newblock Stability selection.
\newblock \emph{J. R. Stat. Soc., B: Stat. Methodol.}, 72\penalty0
  (4):\penalty0 417--473, 2010.

\bibitem[Mishra et~al.(2022)Mishra, Malik, Hachiya, J{\"u}rgenson, Namba,
  Posner, Kamanu, Koido, Le~Grand, Shi, et~al.]{mishra2022stroke}
A.~Mishra, R.~Malik, T.~Hachiya, T.~J{\"u}rgenson, S.~Namba, D.~C. Posner,
  F.~K. Kamanu, M.~Koido, Q.~Le~Grand, M.~Shi, et~al.
\newblock Stroke genetics informs drug discovery and risk prediction across
  ancestries.
\newblock \emph{Nature}, 611\penalty0 (7934):\penalty0 115--123, 2022.

\bibitem[Morrison et~al.(2020)Morrison, Knoblauch, Marcus, Stephens, and
  He]{morrison2020mendelian}
J.~Morrison, N.~Knoblauch, J.~H. Marcus, M.~Stephens, and X.~He.
\newblock Mendelian randomization accounting for correlated and uncorrelated
  pleiotropic effects using genome-wide summary statistics.
\newblock \emph{Nat. Genet.}, 52\penalty0 (7):\penalty0 740--747, 2020.

\bibitem[Nam et~al.(2022)Nam, Kim, and Lee]{nam2022genome}
K.~Nam, J.~Kim, and S.~Lee.
\newblock Genome-wide study on 72,298 individuals in korean biobank data for 76
  traits.
\newblock \emph{Cell Genomics}, 2\penalty0 (10), 2022.

\bibitem[Pachter(2014)]{Pachter2014NetworkNonsense}
L.~Pachter.
\newblock The network nonsense of manolis kellis.
\newblock
  \url{https://liorpachter.wordpress.com/2014/02/11/the-network-nonsense-of-manolis-kellis/},
  Feb 2014.

\bibitem[Pazoki et~al.(2021)]{pazoki2021genetic}
R.~Pazoki et~al.
\newblock Genetic analysis in european ancestry individuals identifies 517 loci
  associated with liver enzymes.
\newblock \emph{Nat. Commun.}, 12\penalty0 (1):\penalty0 2579, 2021.

\bibitem[Pearl(2009)]{pearl2009causality}
J.~Pearl.
\newblock \emph{Causality}.
\newblock Cambridge university press, 2009.

\bibitem[Purcell et~al.(2007)]{purcell2007plink}
S.~Purcell et~al.
\newblock Plink: a tool set for whole-genome association and population-based
  linkage analyses.
\newblock \emph{AJHG}, 81\penalty0 (3):\penalty0 559--575, 2007.

\bibitem[Ravikumar et~al.(2010)Ravikumar, Wainwright, and
  Lafferty]{ravikumar2010high}
P.~Ravikumar, M.~J. Wainwright, and J.~D. Lafferty.
\newblock High-dimensional ising model selection using $\ell_1$-regularized
  logistic regression.
\newblock \emph{Ann. Stat.}, pages 1287--1319, 2010.

\bibitem[Ravikumar et~al.(2011)Ravikumar, Wainwright, Raskutti, and
  Yu]{ravikumar2011high}
P.~Ravikumar, M.~J. Wainwright, G.~Raskutti, and B.~Yu.
\newblock High-dimensional covariance estimation by minimizing
  $\ell_1$-penalized log-determinant divergence.
\newblock \emph{Electron. J. Stat.}, 5:\penalty0 935--980, 2011.

\bibitem[Ruan et~al.(2022)Ruan, Lin, Feng, Chen, Lam, Guo, He, Sawa, Martin,
  et~al.]{ruan2022improving}
Y.~Ruan, Y.-F. Lin, Y.-C.~A. Feng, C.-Y. Chen, M.~Lam, Z.~Guo, L.~He, A.~Sawa,
  A.~R. Martin, et~al.
\newblock Improving polygenic prediction in ancestrally diverse populations.
\newblock \emph{Nat. Genet.}, 54\penalty0 (5):\penalty0 573--580, 2022.

\bibitem[Sanderson et~al.(2019)Sanderson, Davey~Smith, Windmeijer, and
  Bowden]{sanderson2019examination}
E.~Sanderson, G.~Davey~Smith, F.~Windmeijer, and J.~Bowden.
\newblock An examination of multivariable mendelian randomization in the
  single-sample and two-sample summary data settings.
\newblock \emph{Int. J. Epidemiol.}, 48\penalty0 (3):\penalty0 713--727, 2019.

\bibitem[Schwarz(1978)]{schwarz1978estimating}
G.~Schwarz.
\newblock Estimating the dimension of a model.
\newblock \emph{Ann. Stat.}, pages 461--464, 1978.

\bibitem[Shi et~al.(2017)Shi, Mancuso, Spendlove, and Pasaniuc]{shi2017local}
H.~Shi, N.~Mancuso, S.~Spendlove, and B.~Pasaniuc.
\newblock Local genetic correlation gives insights into the shared genetic
  architecture of complex traits.
\newblock \emph{AJHG}, 101\penalty0 (5):\penalty0 737--751, 2017.

\bibitem[Sinnott-Armstrong et~al.(2021)]{sinnott2021genetics}
Sinnott-Armstrong et~al.
\newblock Genetics of 35 blood and urine biomarkers in the uk biobank.
\newblock \emph{Nat. Genet.}, 53\penalty0 (2):\penalty0 185--194, 2021.

\bibitem[Stanzick et~al.(2021)]{stanzick2021discovery}
K.~J. Stanzick et~al.
\newblock Discovery and prioritization of variants and genes for kidney
  function in> 1.2 million individuals.
\newblock \emph{Nat. Commun.}, 12\penalty0 (1):\penalty0 4350, 2021.

\bibitem[Sudlow et~al.(2015)]{sudlow2015uk}
C.~Sudlow et~al.
\newblock Uk biobank: an open access resource for identifying the causes of a
  wide range of complex diseases of middle and old age.
\newblock \emph{PLoS Med.}, 12\penalty0 (3):\penalty0 e1001779, 2015.

\bibitem[Surendran et~al.(2020)]{surendran2020discovery}
P.~Surendran et~al.
\newblock Discovery of rare variants associated with blood pressure regulation
  through meta-analysis of 1.3 million individuals.
\newblock \emph{Nat. Genet.}, 52\penalty0 (12):\penalty0 1314--1332, 2020.

\bibitem[Tibshirani(1996)]{tibshirani1996regression}
R.~Tibshirani.
\newblock Regression shrinkage and selection via the lasso.
\newblock \emph{J. R. Stat. Soc. Ser. B Methodol.}, 58\penalty0 (1):\penalty0
  267--288, 1996.

\bibitem[Vershynin(2018)]{vershynin2018high}
R.~Vershynin.
\newblock \emph{High-dimensional probability: An introduction with applications
  in data science}, volume~47.
\newblock Cambridge University Press, 2018.

\bibitem[Vujkovic et~al.(2020)]{vujkovic2020discovery}
M.~Vujkovic et~al.
\newblock Discovery of 318 new risk loci for type 2 diabetes and related
  vascular outcomes among 1.4 million participants in a multi-ancestry
  meta-analysis.
\newblock \emph{Nat. Genet.}, 52\penalty0 (7):\penalty0 680--691, 2020.

\bibitem[Wang and Li(2022)]{wang2022estimation}
J.~Wang and H.~Li.
\newblock Estimation of genetic correlation with summary association
  statistics.
\newblock \emph{Biometrika}, 109\penalty0 (2):\penalty0 421--438, 2022.

\bibitem[Wang et~al.(2022)]{wang2022mendelian}
K.~Wang et~al.
\newblock Mendelian randomization analysis of 37 clinical factors and coronary
  artery disease in east asian and european populations.
\newblock \emph{Genome Med.}, 14\penalty0 (1):\penalty0 1--15, 2022.

\bibitem[Wang and Huang(2014)]{wang2014review}
Y.~R. Wang and H.~Huang.
\newblock Review on statistical methods for gene network reconstruction using
  expression data.
\newblock \emph{J. Theor. Biol.}, 362:\penalty0 53--61, 2014.

\bibitem[Welsh et~al.(2010)]{welsh2010unraveling}
P.~Welsh et~al.
\newblock Unraveling the directional link between adiposity and inflammation: a
  bidirectional mendelian randomization approach.
\newblock \emph{J. Clin. Endocrinol. Metab.}, 95\penalty0 (1):\penalty0 93--99,
  2010.

\bibitem[Yan et~al.(2020)Yan, Liang, Gao, Wang, Fujioka, Zhu, and
  Wang]{yan2020fam222a}
T.~Yan, J.~Liang, J.~Gao, L.~Wang, H.~Fujioka, X.~Zhu, and X.~Wang.
\newblock Fam222a encodes a protein which accumulates in plaques in
  alzheimer’s disease.
\newblock \emph{Nat. Commun.}, 11\penalty0 (1):\penalty0 411, 2020.

\bibitem[Yang et~al.(2010)]{yang2010common}
J.~Yang et~al.
\newblock Common snps explain a large proportion of the heritability for human
  height.
\newblock \emph{Nat. Genet.}, 42\penalty0 (7):\penalty0 565--569, 2010.

\bibitem[Yang et~al.(2021)Yang, Zhou, and Pan]{yang2021estimation}
Y.~Yang, J.~Zhou, and J.~Pan.
\newblock Estimation and optimal structure selection of high-dimensional
  toeplitz covariance matrix.
\newblock \emph{J. Multivar. Anal.}, 184:\penalty0 104739, 2021.

\bibitem[Ye et~al.(2021)Ye, Shao, and Kang]{ye2021debiased}
T.~Ye, J.~Shao, and H.~Kang.
\newblock Debiased inverse-variance weighted estimator in two-sample
  summary-data mendelian randomization.
\newblock \emph{Ann. Stat.}, 49\penalty0 (4):\penalty0 2079--2100, 2021.

\bibitem[Yi(2017)]{yi2017statistical}
G.~Y. Yi.
\newblock \emph{Statistical analysis with measurement error or
  misclassification: strategy, method and application}.
\newblock Springer, 2017.

\bibitem[Yuan and Lin(2007)]{yuan2007model}
M.~Yuan and Y.~Lin.
\newblock Model selection and estimation in the gaussian graphical model.
\newblock \emph{Biometrika}, 94\penalty0 (1):\penalty0 19--35, 2007.

\bibitem[Zhang(2010)]{zhang2010nearly}
C.-H. Zhang.
\newblock Nearly unbiased variable selection under minimax concave penalty.
\newblock \emph{Ann. Stat.}, pages 894--942, 2010.

\bibitem[Zhang and Zou(2014)]{zhang2014sparse}
T.~Zhang and H.~Zou.
\newblock Sparse precision matrix estimation via lasso penalized d-trace loss.
\newblock \emph{Biometrika}, 101\penalty0 (1):\penalty0 103--120, 2014.

\bibitem[Zhao and Zhu(2022)]{zhao2022genetic}
B.~Zhao and H.~Zhu.
\newblock On genetic correlation estimation with summary statistics from
  genome-wide association studies.
\newblock \emph{J. Am. Stat. Assoc.}, 117\penalty0 (537):\penalty0 1--11, 2022.

\bibitem[Zheng et~al.(2018)Zheng, Aragam, Ravikumar, and Xing]{zheng2018dags}
X.~Zheng, B.~Aragam, P.~K. Ravikumar, and E.~P. Xing.
\newblock Dags with no tears: Continuous optimization for structure learning.
\newblock \emph{Adv. Neural Inf. Process Syst.}, 31, 2018.

\bibitem[Zhu et~al.(2021)Zhu, Li, Xu, and Wang]{zhu2021iterative}
X.~Zhu, X.~Li, R.~Xu, and T.~Wang.
\newblock An iterative approach to detect pleiotropy and perform mendelian
  randomization analysis using gwas summary statistics.
\newblock \emph{Bioinformatics}, 37\penalty0 (10):\penalty0 1390--1400, 2021.

\bibitem[Zhu et~al.(2023)Zhu, Yang, Lorincz-Comi, Li, Bentley, de~Vries, Brown,
  Morrison, Rotimi, Gauderman, et~al.]{zhu2023new}
X.~Zhu, Y.~Yang, N.~Lorincz-Comi, G.~Li, A.~Bentley, P.~S. de~Vries, M.~Brown,
  A.~C. Morrison, C.~Rotimi, W.~J. Gauderman, et~al.
\newblock A new approach to identify gene-environment interactions and reveal
  new biological insight in complex traits.
\newblock \emph{Research Square}, pages rs--3, 2023.

\bibitem[Zhu et~al.(2015)]{zhu2015meta}
X.~Zhu et~al.
\newblock Meta-analysis of correlated traits via summary statistics from gwass
  with an application in hypertension.
\newblock \emph{AJHG}, 96\penalty0 (1):\penalty0 21--36, 2015.

\end{thebibliography}
\end{document}